# Security and Privacy Issues in Cloud Computing

**Jaydip Sen**
*Innovation Labs, Tata Consultancy Services Ltd., Kolkata, INDIA*

## ABSTRACT

Cloud computing transforms the way *information technology* (IT) is consumed and managed, promising improved cost efficiencies, accelerated innovation, faster time-to-market, and the ability to scale applications on demand (Leighton, 2009). According to Gartner, while the hype grew exponentially during 2008 and continued since, it is clear that there is a major shift towards the cloud computing model and that the benefits may be substantial (Gartner Hype-Cycle, 2012). However, as the shape of the cloud computing is emerging and developing rapidly both conceptually and in reality, the legal/contractual, economic, service quality, interoperability, security and privacy issues still pose significant challenges. In this chapter, we describe various service and deployment models of cloud computing and identify major challenges. In particular, we discuss three critical challenges: regulatory, security and privacy issues in cloud computing. Some solutions to mitigate these challenges are also proposed along with a brief presentation on the future trends in cloud computing deployment.

## INTRODUCTION

As per the definition provided by the National Institute for Standards and Technology (NIST) (Badger et al., 2011), "*cloud computing* is a model for enabling convenient, on-demand network access to a shared pool of configurable computing resources (e.g., networks, servers, storage, applications, and services) that can be rapidly provisioned and released with minimal management effort or service provider interaction". It represents a paradigm shift in information technology many of us are likely to see in our lifetime. While the customers are excited by the opportunities to reduce the capital costs, and the chance to divest themselves of infrastructure management and focus on core competencies, and above all the agility offered by the on-demand provisioning of computing, there are issues and challenges which need to be addressed before a ubiquitous adoption may happen.

Cloud computing refers to both the applications delivered as services over the Internet and the hardware and systems software in the datacenters that provide those services. There are four basic cloud delivery models, as outlined by NIST (Badger et al., 2011), based on who provides the cloud services. The agencies may employ one model or a combination of different models for efficient and optimized delivery of applications and business services. These four delivery models are: (i) *Private cloud* in which cloud services are provided solely for an organization and are managed by the organization or a third party. These services may exist off-site. (ii) *Public cloud* in which cloud services are available to the public and owned by an organization selling the cloud services, for example, Amazon cloud service. (iii) *Community cloud* in which cloud services are shared by several organizations for supporting a specific community that has shared concerns (e.g., mission, security requirements, policy, and compliance considerations). These services may be managed by the organizations or a third party and may exist off-site. A special case of community cloud is the Government or G-Cloud. This type of cloud computing is provided by one or more agencies (service provider role), for use by all, or most, government agencies (user role). (iv) *Hybrid cloud* which is a composition of different cloud computing infrastructure (public, private or community). An example for hybrid cloud is the data stored in private cloud of a travel agency that is manipulated by a program running in the public cloud.

From the perspective of service delivery, NIST has identified three basic types of cloud service offerings. These models are: (i) *Software as a service* (SaaS) which offers renting application functionality from a service provider rather than buying, installing and running software by the user. (ii) *Platform as a service* (PaaS) which provides a platform in the cloud, upon which applications can be developed and executed. (iii) *Infrastructure as a service* (IaaS) in which the vendors offer computing power and storage space on demand.

From a hardware point of view, three aspects are new in the paradigm of cloud computing (Armbrust et al., 2009). These aspects of cloud computing are: (i) The illusion of infinite computing resources available on demand, thereby eliminating the need for cloud computing users to plan far ahead for provisioning. (ii) The elimination of an up-front commitment by cloud users, thereby allowing companies to start small and increase hardware resources only when there is an increase in their needs. (iii) The ability to pay for use of computing resources on a short-term basis as needed and release them when the resources are not needed, thereby rewarding conservation by *letting machines and storage go when they are no longer useful*. In a nutshell, cloud computing has enabled operations of large-scale data centers which has led to significant decrease in operational costs of those data centers. On the consumer side, there are some obvious benefits provided by cloud computing. A painful reality of running IT services is the fact that in most of the times, peak demand is significantly higher than the average demand. The resultant massive over-provisioning that the companies usually do is extremely capital-intensive and wasteful. Cloud computing has allowed and will allow even more seamless scaling of resources as the demand changes.

In spite of the several advantages that cloud computing brings along with it, there are several concerns and issues which need to be solved before ubiquitous adoption of this computing paradigm happens. First, in cloud computing, the user may not have the kind of control over his/her data or the performance of his/her applications that he/she may need, or the ability to audit or change the processes and policies under which he/she must work. Different parts of an application might be in different place in the cloud that can have an adverse impact on the performance of the application. Complying with regulations may be difficult especially when talking about cross-border issues – it should also be noted that regulations still need to be developed to take all aspects of cloud computing into account. It is quite natural that monitoring and maintenance is not as simple a task as compared to what it is for PCs sitting in the Intranet. Second, the cloud customers may risk losing data by having them locked into proprietary formats and may lose control over their data since the tools for monitoring who is using them or who can view them are not always provided to the customers. Data loss is, therefore, a potentially real risk in some specific deployments. Third, it may not be easy to tailor *service-level agreements* (SLAs) to the specific needs of a business. Compensation for downtime may be inadequate and SLAs are unlikely to cover the concomitant damages. It is sensible to balance the cost of guaranteeing internal uptime against the advantages of opting for the cloud. Fourth, leveraging cost advantages may not always be possible always. From the perspective of the organizations, having little or no capital investment may actually have tax disadvantages. Finally, the standards are immature and insufficient for handling the rapidly changing and evolving technologies of cloud computing. Therefore, one cannot just move applications to the cloud and expect them to run efficiently. Finally, there are latency and performance issues since the Internet connections and the network links may add to latency or may put constraint on the available bandwidth.

## ARCHICTECTURE OF CLOUD COMPUTING

In this section, we present a top-level architecture of cloud computing that depicts various cloud service delivery models. Cloud computing enhances collaboration, agility, scale, availability and provides the potential for cost reduction through optimized and efficient computing. More specifically, cloud describes the use of a collection of distributed services, applications, information and infrastructure comprised of pools of compute, network, information and storage resources (CSA Security Guidance, 2009). These components can be rapidly orchestrated, provisioned, implemented and decommissioned using an on-demand utility-like model of allocation and consumption. Cloud services are most often, but not always, utilized in conjunction with an enabled by virtualization technologies to provide dynamic integration, provisioning, orchestration, mobility and scale.

While the very definition of cloud suggests the decoupling of resources from the physical affinity to and location of the infrastructure that delivers them, many descriptions of cloud go to one extreme or another by either exaggerating or artificially limiting the many attributes of cloud. This is often purposely done in an attempt to inflate or marginalize its scope. Some examples include the suggestions that for a service to be cloud-based, that the Internet must be used as a transport, a web browser must be used as an access modality or that the resources are always shared in a multi-tenant environment outside of the "perimeter." What is missing in these definitions is context.

From an architectural perspective, given this abstracted evolution of technology, there is much confusion surrounding how cloud is both similar and different from existing models and how these similarities and differences might impact the organizational, operational and technological approaches to cloud adoption as it relates to traditional network and information security practices. There are those who say cloud is a novel sea-change and technical revolution while other suggests it is a natural evolution and coalescence of technology, economy and culture. The real truth is somewhere in between.

There are many models available today which attempt to address cloud from the perspective of academicians, architects, engineers, developers, managers and even consumers. The architecture that we will focus on this chapter is specifically tailored to the unique perspectives of IT network deployment and service delivery.

Cloud services are based upon five principal characteristics that demonstrate their relation to, and differences from, traditional computing approaches (CSA Security Guidance, 2009). These characteristics are: (i) abstraction of infrastructure, (ii) resource democratization, (iii) service oriented architecture, (iv) elasticity/dynamism, (v) utility model of consumption and allocation.

**Abstraction of infrastructure:** The computation, network and storage infrastructure resources are abstracted from the application and information resources as a function of service delivery. Where and by what physical resource that data is processed, transmitted and stored on becomes largely opaque from the perspective of an application or services' ability to deliver it. Infrastructure resources are generally pooled in order to deliver service regardless of the tenancy model employed – shared or dedicated. This abstraction is generally provided by means of high levels of virtualization at the chipset and operating system levels or enabled at the higher levels by heavily customized file systems, operating systems or communication protocols.

**Resource democratization:** The abstraction of infrastructure yields the notion of resource democratization- whether infrastructure, applications, or information – and provides the capability for pooled resources to be made available and accessible to anyone or anything authorized to utilize them using standardized methods for doing so.

**Service-oriented architecture:** As the abstraction of infrastructure from application and information yields well-defined and loosely-coupled resource democratization, the notion of utilizing these components in whole or part, alone or with integration, provides a services oriented architecture where resources may be accessed and utilized in a standard way. In this model, the focus is on the delivery of service and not the management of infrastructure.

**Elasticity/dynamism:** The on-demand model of cloud provisioning coupled with high levels of automation, virtualization, and ubiquitous, reliable and high-speed connectivity provides for the capability to rapidly expand or contract resource allocation to service definition and requirements using a self-service model that scales to as-needed capacity. Since resources are pooled, better utilization and service levels can be achieved.

**Utility model of consumption and allocation:** The abstracted, democratized, service-oriented and elastic nature of cloud combined with tight automation, orchestration, provisioning and self-service then allows for dynamic allocation of resources based on any number of governing input parameters. Given the visibility at an atomic level, the consumption of resources can then be used to provide a metered utility-cost and usage model. This facilitates greater cost efficacies and scale as well as manageable and predictive costs.

## Cloud Service Delivery Models

Three archetypal models and the derivative combinations thereof generally describe cloud service delivery. The three individual models are often referred to as the "SPI MODEL", where "SPI" refers to Software, Platform and Infrastructure (as a service) respectively (CSA Security Guidance, 2009).

**Software as a Service (SaaS):** The capability provided to the consumer is to use the provider's applications running on a cloud infrastructure and accessible from various client devices through a thin client interface such as web browser. In other words, in this model, a complete application is offered to the customer as a service on demand. A single instance of the service runs on the cloud and multiple end users are services. On the customers' side, there is no need for upfront investment in servers or software licenses, while for the provider, the costs are lowered, since only a single application needs to be hosted and maintained. In summary, in this model, the customers do not manage or control the underlying cloud infrastructure, network, servers, operating systems, storage, or even individual application capabilities, with the possible exception of limited user-specific application configuration settings. Currently, SaaS is offered by companies such as Google, Salesforce, Microsoft, Zoho etc.

**Platform as a Service (PaaS):** In this model, a layer of software or development environment is encapsulated and offered as a service, upon which other higher levels of service are built. The customer has the freedom to build his own applications, which run on the provider's infrastructure. Hence, a capability is provided to the customer to deploy onto the cloud infrastructure customer-created applications using programming languages and tools supported by the provider (e.g., Java, Python, .Net etc.). Although the customer does not manage or control the underlying cloud infrastructure, network, servers, operating systems, or storage, but he/she has the control over the deployed applications and possibly over the application hosting environment configurations. To meet manageability and scalability requirements of the applications, PaaS providers offer a predefined combination of operating systems and application servers, such as LAMP (Linux, Apache, MySql and PHP) platform, restricted J2EE, Ruby etc. Some examples of PaaS are: Google's App Engine, Force.com, etc.

**Infrastructure as a Service (IaaS):** This model provides basic storage and computing capabilities as standardized services over the network. Servers, storage systems, networking equipment, data center space etc. are pooled and made available to handle workloads. The capability provided to the customer is to rent processing, storage, networks, and other fundamental computing resources where the customer is able to deploy and run arbitrary software, which can include operating systems and applications. The customer does not manage or control the underlying cloud infrastructure but has the control over operating systems, storage, deployed applications, and possibly select networking components (e.g., firewalls, load balancers etc.). Some examples of IaaS are: Amazon, GoGrid, 3 Tera etc.

Understanding the relationship and dependencies between these models is critical. IaaS is the foundation of all cloud services with PaaS building upon IaaS, and SaaS-in turn – building upon PaaS. An architecture of cloud layer model is depicted in Figure 1.

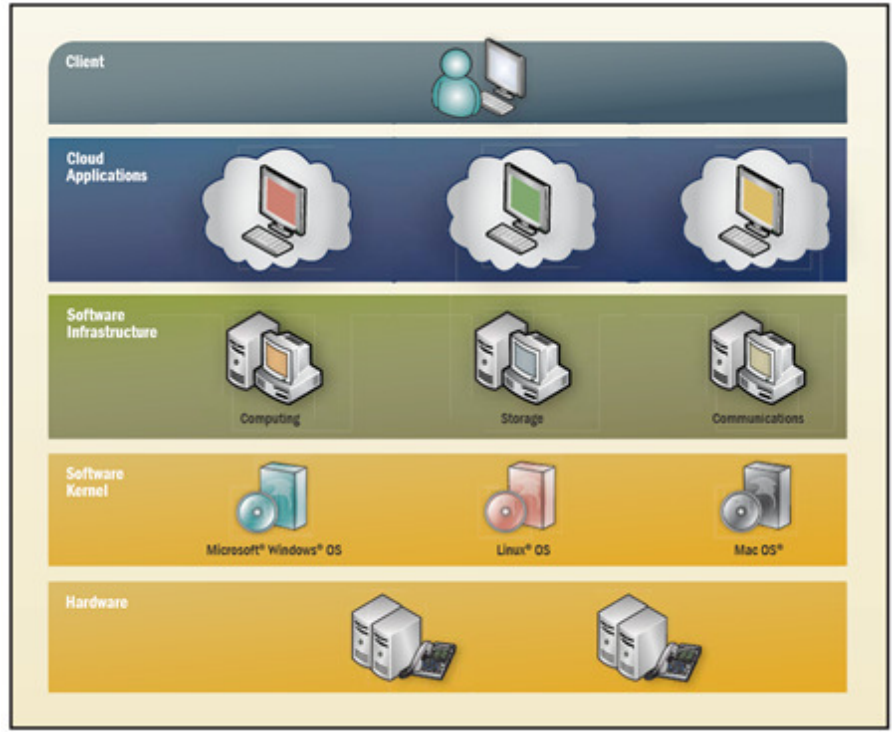

*Figure 1: An architecture of the layer model of cloud computing*

## Cloud Service Deployment and Consumption Models

Regardless of the delivery model utilized (SaaS, PaaS, IaaS) there are four primary ways in which cloud services are deployed (CSA Security Guidance, 2009). Cloud integrators can play a vital role in determining the right cloud path for a specific organization.

**Public cloud:** Public clouds are provided by a designated service provider and may offer either a single-tenant (dedicated) or multi-tenant (shared) operating environment with all the benefits and functionality of elasticity and the accountability/utility model of cloud. The physical infrastructure is generally owned by and managed by the designated service provider and located within the provider's data centers (off-premises). All customers share the same infrastructure pool with limited configuration, security protections, and availability variances. One of the advantages of a public cloud is that they may be larger than an enterprise cloud, and hence they provide the ability to scale seamlessly on demand.

**Private cloud**: Private clouds are provided by an organization or their designated services and offer a single-tenant (dedicated) operating environment with all the benefits and functionality of elasticity and accountability/utility model of cloud. The private clouds aim to address concerns on data security and offer greater control, which is typically lacking in a public cloud. There are two variants of private clouds: (i) on-premise private clouds and (ii) externally hosted private clouds. The on-premise private clouds, also known as internal clouds are hosted within one's own data center. This model provides a more standardized process and protection, but is limited in aspects of size and scalability. IT departments would also need to incur the capital and operational costs for the physical resources. This is best suited for applications which require complete control and configurability of the infrastructure and security. As the name implies, the externally hosted private clouds are hosted externally with a cloud provider in which the provider

**Hybrid cloud:** Hybrid clouds are a combination of public and private cloud offerings that allow for transitive information exchange and possibly application compatibility and portability across disparate cloud service offerings and providers utilizing standard or proprietary methodologies regardless of ownership or location. With a hybrid cloud, service providers can utilize third party cloud providers in a full or partial manner, thereby increasing the flexibility of computing. The hybrid cloud model is capable of providing on-demand, externally provisioned scale. The ability to augment a private cloud with the resources of a public cloud can be used to manage any unexpected surges in workload.

Table 1: Summary of the various features of cloud deployment models

| Deployment Model | Managed By | Infrastructure Owned By | Infrastructure Located At | Accessible and Consumed By |
|---|---|---|---|---|
| Public | Third party provider | Third party provider | Off-premise | Untrusted |
| Private | Organization | Organization | On-premise Off-premise | Trusted |
| | Third party provider | Third party provider | On-premise Off-premise | |
| Managed | Third party provider | Third party provider | On-premise | Trusted or Untrusted |
| Hybrid | Both organization and third party provider | Both organization and third party provider | Both on-premise and off-premise | Trusted or Untrusted |

**Managed cloud:** Managed clouds are provided by a designated service provider and may offer either a single-tenant (dedicated) or multi-tenant (shared) operating environment with all the benefits and functionality of elasticity and the accountability/utility model of cloud. The physical infrastructure is owned by and/or physically located in the organizations' data centers with an extension of management and security control planes controlled by the designated service provider.

The notion of public, private, managed and hybrid when describing cloud services really denotes the attribution of management and the availability of service to specific consumers of the services. Table 1 summarizes various features of the four cloud deployment models.

When assessing the impact a particular cloud service may have on one's security posture and overall security architecture, it is necessary to classify the assets/resource/service within the context of not only its location but also its criticality and business impact as it relates to management and security. This means that an appropriate level of risk assessment is performed prior to entrusting it to the vagaries of the cloud (CSA Security Guidance, 2009). In addition, it is important to understand various tradeoffs between the various cloud service models:

- Generally, SaaS provides a large amount of integrated features built directly into the offering with the least amount of extensibility and in general a high level of security (or at least a responsibility for security on the part of the service provider).
- PaaS offers less integrated features since it is designed to enable developers to build their own applications on top of the platform, and it is, therefore, more extensible than SaaS by nature. However, this extensibility features trade-offs on security features and capabilities.
- IaaS provides few, if any, application-like features, and provides for enormous extensibility but generally less security capabilities and functionalities beyond protecting the infrastructure itself, since it expects operating systems, applications and contents to be managed and secured by the customers.

In summary, form security perspective, in the three service models of cloud computing, the lower down the stack the cloud service provider stops, the more security capabilities and management the customer is responsible for implementing and managing themselves.

## CLOUD COMPUTING SECURITY AND PRIVACY ISSUES

This section addresses the core theme of this chapter, i.e., the security and privacy-related challenges in cloud computing. There are numerous security issues for cloud computing as it encompasses many technologies including networks, databases, operating systems, virtualization, resource scheduling, transaction management, load balancing, concurrency control and memory management. Therefore, security issues for many of these systems and technologies are applicable to cloud computing. For example, the network that interconnects the systems in a cloud has to be secure. Furthermore, virtualization paradigm in cloud computing leads to several security concerns. For example, mapping the virtual machines to the physical machines has to be carried out securely. Data security involves encrypting the data as well as ensuring that appropriate policies are enforced for data sharing. In addition, resource allocation and memory management algorithms have to be secure. Finally, data mining techniques may be applicable for malware detection in the clouds – an approach which is usually adopted in *intrusion detection systems* (IDSs) (Sen & Sengupta, 2005; Sen et al., 2006b; Sen et al., 2008; Sen, 2010a; Sen, 2010b; Sen 2010c).

As shown in Figure 2, there are six specific areas of the cloud computing environment where equipment and software require substantial security attention (Trusted Computing Group's White Paper, 2010). These six areas are: (1) security of data at rest, (2) security of data in transit, (3) authentication of users/applications/ processes, (4) robust separation between data belonging to different customers, (5) cloud legal and regulatory issues, and (6) incident response.

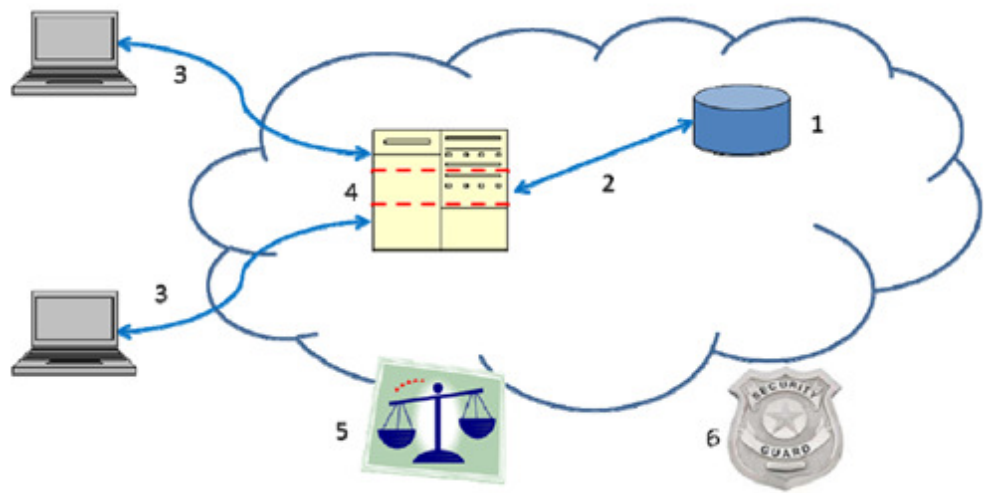

*Figure 2: Areas for security concerns in cloud computing: (1) data at rest, (2) data in transit, (3) authentication, (4) separation between customers, (5) cloud legal and regulatory issues and (6) incident response.*

For securing data at rest, cryptographic encryption mechanisms are certainly the best options. The hard drive manufacturers are now shipping self-encrypting drives that implement trusted storage standards of the trusted computing group (Trusted Computing Group's White Paper, 2010). These self-encrypting drives build encryption hardware into the drive, providing automated encryption with minimal cost or performance impact. Although software encryption can also be used for protecting data, it makes the process slower and less secure since it may be possible for an adversary to steal the encryption key from the machine without being detected.

Encryption is the best option for securing data in transit as well. In addition, authentication and integrity protection mechanisms ensure that data only goes where the customer wants it to go and it is not modified in transit.

Strong authentication is a mandatory requirement for any cloud deployment. User authentication is the primary basis for access control. In the cloud environment, authentication and access control are more important than ever since the cloud and all of its data are accessible to anyone over the Internet. The *trusted computing group's* (TCG's) IF-MAP standard allows for real-time communication between a cloud service provider and the customer about authorized users and other security issues. When a user's access privilege is revoked or reassigned, the customer's identity management system can notify the cloud provider in real-time so that the user's cloud access can be modified or revoked within a very short span of time.

One of the more obvious cloud concerns is separation between a cloud provider's users (who may be competing companies or even hackers) to avoid inadvertent or intentional access to sensitive information. Typically a cloud provider would use *virtual machines* (VMs) and a hypervisor to separate customers. Technologies are currently available that can provide significant security improvements for VMs and virtual network separation. In addition, the *trusted platform module* (TPM) can provide hardware-based verification of hypervisor and VM integrity and thereby ensure strong network separation and security.

Legal and regulatory issues are extremely important in cloud computing that have security implications. To verify that a cloud provider has strong policies and practices that address legal and regulatory issues, each customer must have its legal and regulatory experts inspect cloud provider's policies and practices to ensure their adequacy. The issues to be considered in this regard include data security and export, compliance, auditing, data retention and destruction, and legal discovery. In the areas of data retention and deletion, trusted storage and trusted platform module access techniques can play a key role in limiting access to sensitive and critical data.

As part of expecting the unexpected, customers need to plan for the possibility of cloud provider security breaches or user misbehavior. An automated response o at least automated notification is the best solution for this purpose. The IF-MAP (Metadata Access protocol) of the *trusted computing group* (TCG) specification enables the integration of different security systems and provides real-time notifications of incidents and of user misbehavior.

## Threat Vectors- What to Worry About in Security

How does the landscape of threats to security and privacy change as organizations shift to cloud-based systems, storage and applications? New vectors are introduced, and old ones can be exploited in new ways. In the following, we briefly discuss some of the threats, highlighting what is genuinely different and new in a world of cloud hosting, what threats are similar to the dominant model of local applications and in-house IT management but will manifest in different ways.

Before categorizing new threats, it is important to acknowledge that the structure of many cloud architectures can mitigate or negate some current security threats. If data are kept in the cloud, for example, then a lost or stolen laptop is much less likely to put sensitive information at risk. Standardized interfaces could make security management easier (ENISA, 2009), while the scale of a provider hosting many parties can generate more information for better threat monitoring. Centralized security management and monitoring can be more effective than local efforts by IT professionals with limited security experience.

Still, moving critical systems and data to a network-accessible framework introduces new classes of vulnerabilities in and of itself, by creating new surfaces to attack and new interfaces to exploit. When those network resources are built on systems, platforms and applications shared with others, another set of threat vectors is introduced. The control mechanisms itself can be attacked, breaking down isolation between users, potentially allowing another user to access data or resources. Even without direct access, a providers' other clients can learn valuable transaction data about an organization (Ristenpart et al., 2009). The shared architecture also puts a cloud user at risk from other cloud users if their bad behavior draws attention from either law enforcement or media, leading to hardware seizure or bad publicity (Molnar & Schechter, 2010).

Some threat vectors are not new to cloud, but have somewhat different dynamics. In classic IT architecture, PCs inside the organization may be at risk of compromise through a host of attack vectors exploiting local applications such as browses or documents viewers. If less data is stored locally, less is immediately at risk, but now the attacker could compromise credentials to gain access to the user's cloud privileges. A compromise to an entire Gmail database probably began with a compromised PC (Zetter, 2010). Similarly, in an attack on the Twitter management team in 2009, a compromised email password led to exposure of a wide range of other important documents in other cloud infrastructures (Lowensohn & McCarthy, 2009). Shared authentication tokens can lead to brittle defenses.

Organizations must be careful to safeguard data as they move it around their organization, even without the benefit of cloud computing. When they no longer need data, it must be properly deleted, or else risk leaking sensitive data to the outside (Garfinkel & Shelat, 2003). When relying on a cloud service to handle data, appropriate care must be made to arrange for appropriate security management practices, such as encryption and appropriate deletion.

Similarly, all organizations are vulnerable to an insider attack from a trusted insider, but moving things to the cloud can raise the costs of misplaced trust. A cloud system with a well-thought out identity interface and a clear access control system can restrict access and foster accountability. However, a unified data system with more people accessing more different types of data through more applications can actually make it harder to appropriately limit access and detect misuse (Sinclair & Smith, 2008).

## Security Issues in Cloud Computing

Security in the cloud is achieved, in part, through third party controls and assurance much like in traditional outsourcing arrangements. But since there is no common cloud computing security standard, there are additional challenges associated with this. Many cloud vendors implement their own proprietary standards and security technologies, and implement differing security models, which need to be evaluated on their own merits. In a vendor cloud model, it is ultimately down to adopting customer organizations to ensure that security in the cloud meets their own security polices through requirements gathering provider risk assessments, due diligence, and assurance activities (CPNI Security Briefing, 2010).

Thus, the security challenges faced by organizations wishing to use cloud services are not radically different from those dependent on their own in-house managed enterprises. The same internal and external threats are present and require risk mitigation or risk acceptance. In the following, we examine the information security challenges that adopting organizations will need to consider, either through assurance activities on the vendor or public cloud providers or directly, through designing and implementing security control in a privately owned cloud. In particular, we examine the following issues:

- The treats against information assets residing in cloud computing environments.
- The types of attackers and their capability of attacking the cloud.
- The security risks associated with the cloud, and where relevant considerations of attacks and countermeasures.
- Emerging cloud security risks.
- Some example cloud security incidents.

## Cloud Security Threats

The threats to information assets residing in the cloud can vary according to the cloud delivery models used by cloud user organizations. There are several types of security threats to which cloud computing is vulnerable. Table 2 provides an overview of the threats for cloud customers categorized according to the *confidentiality, integrity* and *availability* (CIA) security model and their relevance to each of the cloud service delivery model.

Table 2: A list of cloud security threats

| Threat | Description |
|---|---|
| **Confidentiality** | |
| Insider user threats:<br>• Malicious cloud provider user<br>• Malicious cloud customer user<br>• Malicious third party user (Supporting either the cloud provider or customer organizations) | The threat of insiders accessing customer data held within the cloud is greater as each of the delivery models can introduce the need for multiple internal users:<br>SaaS – cloud customer and provider administrators<br>PaaS- application developers and test environment managers<br>IaaS- third party platform consultants |
| External attacker threats:<br>• Remote software attack of cloud infrastructure<br>• Remote software attack of cloud applications<br>• Remote hardware attack against the cloud<br>• Remote software and hardware attack against cloud user organizations' endpoint software and hardware<br>• Social engineering of cloud provider users, and cloud customer users. | The threat from external attackers may be perceived to apply more to public Internet facing clouds, however all types of cloud delivery models are affected by external attackers, particularly in private clouds where user endpoints can be targeted. Cloud providers with large data stores holding credit card details, personal information and sensitive government or intellectual property, will be subjected to attacks from groups, with significant resources, attempting to retrieve data. This includes the threat of hardware attack, social engineering and supply chain attacks by dedicated attackers. |
| Data leakage:<br>• Failure of security access rights across multiple domains<br>• Failure of electronic and physical transport systems for cloud data and backups | A threat from widespread data leakage amongst many, potentially competitor organizations, using the same cloud provider could be caused by human error or faulty hardware that will lead to information compromise. |
| **Integrity** | |
| Data segregation:<br>• Incorrectly defined security perimeters<br>• Incorrect configuration of virtual machines and hypervisors | The integrity of data within complex cloud hosting environments such as SaaS configured to share computing resource amongst customers could provide a threat against data integrity if system resources are effectively segregated. |
| User access:<br>• Poor identity and access management procedures | Implementation of poor access control procedures creates many threat opportunities, for example that disgruntled ex-employees of cloud provider organizations maintain remote access to administer customer cloud services, and can cause intentional damage to their data sources. |
| Data quality:<br>• Introduction of faulty application or infrastructure components | The threat of impact of data quality is increased as cloud providers host many customers' data. The introduction of a faulty or misconfigured component required by another cloud user could potentially impact the integrity of data for other cloud users sharing infrastructure. |
| **Availability** | |
| Change management:<br>• Customer penetration testing impacting other cloud customers<br>• Infrastructure changes upon cloud provider, customer and third party systems impacting cloud customers | As the cloud provider has increasing responsibility for change management within all cloud delivery models, there is a threat that changes could introduce negative effects. These could be caused by software or hardware changes to existing cloud services. |
| Denial of service threat: | The threat of denial of service against available cloud |
| • Network bandwidth distributed denial of service<br>• Network DNS denial of service<br>• Application and data denial of service | computing resource is generally an external threat against public cloud services. However, the threat can impact all cloud service models as external and internal threat agents could introduce application or hardware components that cause a denial of service. |
| Physical disruption:<br>• Disruption of cloud provider IT services through physical access<br>• Disruption of cloud customer IT services through physical access<br>• Disruption of third party WAN providers services | The threat of disruption to cloud services caused by physical access is different between large cloud service providers and their customers. These providers should be experienced in securing large data center facilities and have considered resilience among other availability strategies. There is a threat that cloud user infrastructure can be physically disrupted more easily whether by insiders or externally where less secure office environments or remote working is standard practice. |
| Exploiting weak recovery procedures:<br>• Invocation of inadequate disaster recovery or business continuity processes | The threat of inadequate recovery and incident management procedures being initiated is heightened when cloud users consider recovery of their own in house systems in parallel with those managed by third party cloud service providers. If these procedures are not tested then the impact upon recovery time may be significant. |

## Types of Attackers in Cloud Computing

Many of the security threats and challenges in cloud computing will be familiar to organizations managing in house infrastructure and those involved in traditional outsourcing models. Each of the cloud computing service delivery models' threats result from the attackers that can be divided into two groups as depicted in Table 3.

Table 3: A list of attacks on cloud computing environments

| | |
|---|---|
| **Internal attackers** | An internal attacker has the following characteristics:<br>• Is employed by the cloud service provider, customer or other third party provider organization supporting the operation of a cloud service<br>• May have existing authorized access to cloud services, customer data or supporting infrastructure and applications, depending on their organizational role<br>• Uses existing privileges to gain further access or support third parties in executing attacks against the confidentiality integrity and availability of information within the cloud service. |
| **External attackers** | An external attacker has the following characteristics:<br>• Is not employed by the cloud service provider, customer or other third party provider organization supporting the operation of a cloud service<br>• Has no authorized access to cloud services, customer data or supporting infrastructure and applications<br>• Exploits technical, operational, process and social engineering vulnerabilities to attack a cloud service provider, customer or third party supporting organization to gain further access to propagate attacks against the confidentiality, integrity and availability of information within the cloud service. |

Although internal and external attackers can be clearly differentiated, their capability to execute successful attacks is what differentiates them as a threat to customers and vendors alike.

In the cloud environment, attackers can be categorized into four types: random, weak, strong, and substantial (CPNI Security Briefing, 2010). Each of these categories is based on ability to instigate a successful attack, rather than on the type of threat they present (i.e., criminal, espionage or terrorism):

- **Random**- The most common type of attacker uses simple tools and techniques. The attacker may randomly scan the Internet trying to find vulnerable components. They will deploy well known tools or techniques that should be easily detected.
- **Weak** – Semi-skilled attackers targeting specific servers/cloud providers by customizing existing publicly available tools or specific targets. Their methods are more advanced as they attempt to customize their attacks using available exploit tools.
- **Strong** – Organized, well-financed and skilled groups of attackers with an internal hierarchy specializing in targeting particular applications and users of the cloud. Generally this group will be an organized crime group specializing in large scale attacks.
- **Substantial** – Motivated, strong attackers not easily detected by the organizations they attack, or even by the relevant law enforcement and investigative organizations specializing in eCrime or cyber security. Mitigating this threat requires greater intelligence on attacks and specialist resources in response to detection of an incident or threat.

## Cloud Security Risks

The security risks associated with each cloud delivery model vary and are dependent on a wide range of factors including the sensitivity of information assets, cloud architectures and security control involved in a particular cloud environment. In the following we discuss these risks in a general context, except where a specific reference to the cloud delivery model is made. Table 4 summarizes the security risks relevant in the cloud computing paradigm.

Table 4: A list of security risks in cloud computing

| Risk | Description |
|------|-------------|
| Privileged user access | Cloud providers generally have unlimited access to user data, controls are needed to address the risk of privileged user access leading to compromised customer data. |
| Data location and segregation | Customers may not know where their data is being stored and there may be a risk of data being stored alongside other customers' information. |
| Data disposal | Cloud data deletion and disposal is a risk, particularly where hardware is dynamically issued to customers based on their needs. The risk of data not being deleted from data stores, backups and physical media during de-commissioning is enhanced within the cloud. |
| e-investigations and Protective monitoring | The ability for cloud customers to invoke their own electronic investigations procedures within the cloud can be limited by the delivery model in use, and the access and complexity of the cloud architecture. Customers cannot effectively deploy monitoring systems on infrastructure they do not own; they must rely on the systems in use by the cloud service provider to support investigations. |
| Assuring cloud security | Customers cannot easily assure the security of systems that they do not directly control without using SLAs and having the right to audit security controls within their agreements. |

## Privileged User Access

Once data is stored in the cloud, the provider has access to that data and also controls access to that data by other entities (including other users of the cloud and other third party suppliers). Maintaining confidentiality of data in the cloud and limiting privileged user access can be achieved by at least one of two approaches by the data owner: first, encryption of the data prior to entry into the cloud to separate the ability to store the data from the ability to make use of it; and second, legally enforcing the requirements of the cloud provider through contractual obligations and assurance mechanisms to ensure that confidentiality of the data is maintained to required standards. The cloud provider must have demonstrable security access control policies and technical solutions in place that prevent privilege escalation by standard users, enable auditing of user actions, and support the segregation of duties principle for privileged users in order to prevent and detect malicious insider activity.

Encryption of data prior to entry into the cloud poses two challenges. For encryption of data to be effective means of maintaining data confidentiality, decryption keys must be segregated securely from the cloud environment to ensure that only an authorized party can decrypt data. This could be achieved by storing keys on segregated systems in house or by storing keys with a second provider.

An additional challenge around encryption in the cloud is to prevent manipulations of encrypted data such that plain text, or any other meaningful data, can be recovered and be used to break the cipher. This constraint in encryption technology means that cloud providers must not be granted unlimited ability to store and archive encrypted data. If the cloud user organization permits the cloud service provider to handle unencrypted data, then the cloud service provider must provide assurance that the data will be protected from unauthorized access, both internally and externally. Within the cloud, the generation and use of cryptographic keys for each cloud customer could be used to provide another level of protection above and beyond data segregation controls. However, providers need robust key management processes in place and the challenge for customers then becomes gaining assurances over that process.

A strong or substantial attacker could exploit weak encryption policies, and privileged cloud provider management access, to recover customer data using a complex software or hardware attack on user endpoint devices, or cloud infrastructure devices. This attack may involve long term compromise of the cloud provider supply chain, or social engineering of a particular cloud customer user.

The use of encryption technology may also be subject to limitations or specific requirements depending on the jurisdiction in which the cloud provider will be storing cloud customers' data. For example, in some countries, the use of encryption technologies may be restricted based upon the type of encryption or its purpose of operation. Cloud customers should review whether the application of encryption as mandated by the local jurisdiction of the cloud provider is acceptable and does not enhance risk to their data. For example, in the UK, the Regulatory Investigatory Powers Act (RIPA) can impose a legal obligation to disclose encryption keys to enable access to data by security and law enforcement agencies. Cloud customers should ensure that they understand their obligations within all of the jurisdictions used by the cloud provider, and have policies and procedures in place to deal with specific external enquiries with respect to encrypted data.

## Data Location and Segregation

Data location and data segregation are of particular importance in the cloud, given the disparate physical location of data and shared computing resources. Cloud users may be under statutory, regulatory or contractual obligations to ensure that data is held, processed and managed in a certain way. There are a number of associated security risks in this situation:

- The cloud provider being required to disclose data (and potentially decryption keys) or hand over physical media to a third party or statutory authority.
- Development of liabilities to pay tax to local authorities as a result of processing sales or other transactions within their jurisdictions.
- Environmental hazards such as earthquakes, flooding, and extreme weather affecting the security of customer data, and

- Macro-economic hazards such as hyper-inflation or deep recession affecting the providers' services and personnel conditions.

Central storage arrangements in cloud computing also provide attackers with a far richer target of information. In a single attack, attackers could potentially gain access to confidential information belonging to several customer organizations. If adequate segregation of data isn't applied many customers may find themselves suffering a security breach due to an incident that should have been limited to a single customer.

Virtualization is one of a number of enabling technologies of cloud computing that itself is a run-time method of segregation for processing data. Many of the security concerns and issues associated with virtualization are relevant in cloud computing, regardless of whether or not the cloud service provider employs virtualization technologies. Security of data depends on having adequate security controls in each of the layers of the virtualized environment. In addition, secure deletion of memory and storage must be used to prevent data loss in a multi-tenant environment where systems are reused.

The hypervisor layer between the hardware and virtual machine / guest OS ahs privileged access to layers above. It also has a great deal of control over hardware, and increasingly so, as hardware manufacturers implement hypervisor functions directly into chipsets and CPUs. Cloud users, therefore, need to assess cloud service providers' use and operation of virtualization technologies and whether the risk profile can be tolerated.

## Data Disposal

Cloud services that offer data storage typically provide either guarantees or service-level objectives around high availability of that data. Cloud providers achieve this by keeping multiple copies of the data. Where the cloud customer has a requirement to delete data, cloud-based storage may be inappropriate for that data at all points in its lifecycle.

Depending on the type of data hosted in the cloud, customers may require providers to delete data in accordance with industry standards. Unless the cloud architecture specifically limits the media on which data may be stored and the data owner can mandate use of media sanitization techniques on that media in line with the required standards, customers may need to preclude their data from being transmitted in the cloud.

## e-Investigations and Protective Monitoring

Implementing protective monitoring in the cloud presents challenges for both cloud customers and providers given the disparate location of physical data and the high number of providers involved. While cloud enabling technologies are designed to place a security perimeter between the cloud service systems and the cloud users, vulnerabilities in this layer of security cannot be ruled out altogether. There is a risk of insider threats and attacks on the cloud and this is likely to require expertise in e-investigations and protective monitoring.

Effective protective monitoring of cloud-based information assets is likely to require integration between monitoring tools employed by the cloud provider as well as tools employed by the cloud user. Tracing actions back to accountable users and administrators in the cloud may require an integrated or federated (mutual trust) identity management and associated logging system which permits unambiguous identification of all authorized individual with access to the cloud resources.

Managing identity and access in the cloud for an enterprise is likely to require integration with a pre-existing identity management system. Federating the cloud customer's identity management system with the cloud provider's identity and access management system is one solution.

Protective monitoring of the cloud will, in certain cases, depend on cloud customers' ability to trace actions back to all authorized identities in both cloud and customer IT environment. This is likely to require a federated identity management approach which encompasses the cloud users as well as the cloud service provider.

The technology supporting federated identity management is currently in its very early stages, with several competing standards vying for dominance in a landscape of numerous proprietary identity management technologies.

Access to accurate information is, of course, vital in investigating incidents. Having access to data within protective monitoring logging systems, and the ability to carry out forensic investigations on computing devices and other infrastructure within a cloud environment may be difficult for cloud customers pursuing an investigation. Therefore, customers should address this issue within their contractual agreements with providers, and understand how their provider implements protective monitoring within their cloud environment. Customers placing specific requirements within a contract relating to investigations will need to consider how their investigation team integrates with their equivalent investigations team within the cloud provider organization. This is of particular importance where investigations are taking place on multi-tenant systems and providers have a responsibility to protect other customers' data. Generally, collecting digital evidence within the cloud should be the responsibility of the cloud service provider, and it should be handed over as part of the chain of custody of evidence to the customer for their own investigation process. Should customers request more direct access to specific data devices that are part of a shared customer infrastructure, then the provider may choose to change the architecture of that customer's service which may substantially increase the costs to the customer and may impact the original business case for choosing cloud services.

## Assessing the Security of a Third Party Cloud Provider

One of the most significant challenges for vendor cloud customers in particular is assurance over the security controls of their cloud provider (CPNI Security Briefing, 2010). This is exacerbated by the fact that there is currently no common industry cloud computing security standard from which customers can benchmark their providers. Customers are primarily concerned with the following issues:

- *Defining security requirements* – The customers' information security requirements are derived from the organization's own policy, legal and regulatory obligations, and may carry through from other contracts or SLAs that the company has with its customers.
- *Due diligence on cloud service providers* – Prospective cloud customers should undertake proper due-diligence on providers before entering into a formal relationship. Detailed due-diligence investigations can provide an unbiased and valuable insight into a providers' past track record, including its financial status, legal action taken against the organization and its commercial reputation. Certification schemes such as ISO27001 also provide customers with some assurances that a cloud provider has taken certain steps in its management of information security risks.
- *Managing cloud supplier risks* – The outsourcing of key services to the cloud may require customer organizations to seek new and more mature approaches to risk management and accountability. While cloud computing means that services are outsourced, the risk remains with the customer and it is therefore in the customer's interest to ensure that risks are appropriately managed according to their risk appetite. Effective risk management also requires maturity both in vendor relationship management processes and operational security processes.

## Classification of Security Issues in Cloud Computing

The security issues in cloud computing can be categorized into the following three broad classes:

- Traditional security concerns
- Availability issues
- Third party data control-related issues

In the following, we discuss these three classes of security issues and also highlight some additional security vulnerabilities in cloud computing.

## Traditional Security Issues

These security issues involve computer and network intrusions or attacks that will be made possible or at least easier by moving to the cloud. Cloud providers respond to these concerns by arguing that their security measures and processes are more mature and tested than those of the average company. Another argument, made by the Jericho Forum (Don't Cloud Vision) is: "It could be easier to lock down information if it's administered by a third party rather than in-house, if companies are worried about insider threats… In addition, it may be easier to enforce security via contracts with online services providers than via internal controls."

Concerns in this category include the following:

**VM-level attacks**: VM-level attacks: Potential vulnerabilities in the hypervisor or VM technology used by cloud vendors are a potential problem in multi-tenant architectures. Vulnerabilities have appeared in VMWare (Security Tracker: VMWare Shared Folder Bug), Xen (Xen Vulnerability), and Microsoft's Virtual PC and Virtual Server (Microsoft Security Bulletin MS07-049). Vendors such as Third Brigade mitigate potential VM-level vulnerabilities through monitoring and firewalls.

**Cloud service providers' vulnerabilities:** These could be platform-level, such as an SQL-injection or cross-site scripting vulnerability. For instance, there have been a couple of recent Google Docs vulnerabilities (Microsoft Security Bulletin MS07-049). IBM has repositioned its Rational AppScan tool, which scans for vulnerabilities in web services as a cloud security service (IBM Blue Cloud Initiative).

**Phishing cloud provider:** Phishers and other social engineers have a new attack vector (Salesforce.Com Warns Customers).

**Expanded network attack surface:** The cloud user must protect the infrastructure used to connect and interact with the cloud, a task complicated by the cloud being outside the firewall in many cases. For instance, an example of how the cloud might attack the machine connecting to it has been shown in (Security Evaluation of Grid).

**Authentication and authorization:** The enterprise authentication and authorization framework does not naturally extend into the cloud. How does a company meld its existing framework to include cloud resources? Furthermore, how does an enterprise merge cloud security data (if available) with its own security metrics and policies?

**Forensics in the cloud**: CLOIDIFIN project (Biggs & Vidalis, 2009) summarizes the difficulty of cloud forensic investigations: "Traditional digital forensic methodologies permit investigators to seize equipment and perform detailed analysis on the media and data recovered. The likelihood, therefore, of the data being removed, overwritten, deleted or destroyed by the perpetrator in this case is low. More closely linked to a cloud computing environment would be businesses that own and maintain their own multi-server type infrastructure, though this would be on a far smaller scale in comparison. However, the scale of the cloud and the rate at which data is overwritten is of concern."

## Availability

These concerns center on critical applications and data being available. Well-publicized incidents of cloud outages include Gmail's one-day outage in mid-October 2008 (Extended Gmail Outage), Amazon S3's over seven-hour downtime on July 20, 2008 (Amazon S3 Availability Event, 2008), and FlexiScale's 18-hour outage on October 31, 2008 (Flexiscale Outage). Maintaining the uptime, preventing denial of service attacks (especially at the single-points-of-failure) and ensuring robustness of computational integrity (i.e. the cloud provider is faithfully running an application and giving valid results) are some of the major issues in this category of threats.

## Third Party Data Control

The legal implications of data and applications being held by a third party are complex and not well understood. There is also a potential lack of control and transparency when a third party holds the data. Part of the hype of cloud computing is that the cloud can be implementation-independent, but in reality, regulatory compliance requires transparency into the cloud. Various security and data privacy issues are prompting some companies to build clouds to avoid these issues and yet retain some of the benefits of cloud computing. However, the following concerns need to be addressed properly.

**Due diligence**: If served a subpoena or other legal action, can a cloud user compel the cloud provider to respond in the required time-frame? A related question is the provability of deletion, relevant to an enterprise's retention policy: How can a cloud user be guaranteed that data has been deleted by the cloud provider?

**Auditability:** Audit difficulty is another side effect of the lack of control in the cloud. Is there sufficient transparency in the operations of the cloud provider for auditing purposes? Currently, this transparency is provided by documentation and manual audits. Performing an on-site audit in a distributed and dynamic multi-tenant computing environment spread all over the globe is a major challenge. Certain regulations will require data and operations to remain in certain geographic locations.

**Contractual obligations**: One problem with using another company's infrastructure besides the uncertain alignment of interests is that there might be surprising legal implications. For instance, a passage from Amazon's terms of use is as follows: "Non-assertion" during and after the term of the agreement, with respect to any of the services that you elect to use, you will not assert nor will you authorize, assist, or encourage any third party to assert, against us or any of our customers, end users, vendors, business partners (including third party sellers on websites operated by or on behalf of us), licensors, sub-licensees, or transferees, any patent infringement or other intellectual property infringement claim with respect to such Services." This could be interpreted that after one uses E2C, one cannot file infringement claims against Amazon. It's not clear whether this non-assert would be upheld by the courts, but any uncertainty is bad for business.

**Cloud provider espionage**: The worry of theft of company propriety information by the cloud provider.

**Transitive nature of contracts**:  Another possible concern is that the contracted cloud provider might itself use sub-contractors, over whom the cloud user either has no control or very less control. However, the sub-contractor must be trusted. For example, the online storage service called "Linkup" uses an online storage company called Nirvanix (Loss of Customer Spurs Closure of LinkUp) for its cloud service. Another example is Carbonite (Latest Cloud Storage Hiccups), which is using its hardware providers for faulty equipment causing loss of customer data.

## Emerging Cloud Security Threats

In the following, we discuss some additional security threats that are relevant in cloud computing and are being detected and researched by academia, security organization and both cloud service providers and the cloud customers.

**Side channel attacks:** An emerging concern for cloud delivery models using virtualization platforms is the risk of side channel attacks causing data leakage across co-resident virtual machine instances. This risk is evolving, though currently is considered to be in its infancy, as the virtual machine technologies mature. However, it is possible that attackers who fail to compromise endpoints or penetrate cloud infrastructure from outside the cloud perimeter, may consider this technique - acting as a rogue customer within a shared cloud infrastructure to access other customers' data.

**Denial of service attacks:** Availability is a primary concern to cloud customers and as such it is equally of concern to the service providers who must design solutions to mitigate this threat. Traditionally, denial of service (DoS) has been associated with network layer distributed attacks flooding infrastructure with excessive traffic in order to cause critical components to fail or to consume all available hardware resources (Sen et al., 2006a; Sen, 2011a; Sen, 2011b). Within a multi-tenant cloud infrastructure, there are more specific threats associated with DoS. Some of these threats are: (a) Shared resource consumption – attacks that deprive other customers of system resources such as thread execution time, memory, storage requests and network interfaces can cause a targeted DoS, (b) Virtual machine and hypervisor exploitation – attacks that exploit vulnerabilities in the underlying hypervisor, or operating system hosting a virtual machine instance will allow attackers to cause targeted outages or instability. Attacks using these methods are designed to circumvent traditionally well-defined cloud architecture that has concentrated on securing against external network-based DoS attacks.

**Social networking attacks:** With the increased popularity of business and personal social networking sites the risk of advanced social engineering attack is increased. Cloud computing systems are targeted due to their large customer data stores. The complex set of relationships between cloud providers, customers, suppliers and vendors means that many employees of these organizations will be listed on social networking sites and be connected to each other. Attackers can setup identities to gain trust, and use online information to determine relationships and roles of staff to prepare their attacks. A combination of technical attack and social engineering attacks can be deployed against a target user by taking advantage of the people they know and the online social network they use.

**Mobile device attacks:** The use if smart phones has increased and cloud connectivity is now no longer limited to laptop or desktop computing devices. Attacks are now emerging that are targeted for mobile devices and rely on features traditionally associated with laptops and desktops, including: (i) rich application programming interfaces (APIs) that support network communications and background services, (ii) always on wireless Internet access, and (iii) large local data storage capabilities. As mobile devices now have these equivalent features, Internet-based spyware, worms or even physical attacks may be more likely to occur against mobile devices, as they are potentially a less risky target to an attacker that wishes to remain undetected. This is generally supported by the fact that most mobile devices do not have the equivalent security features enabled, or in some case available. For example, mature antimalware, antivirus or full disk encryption technologies are not widespread on current available smart phones.

**Insider and organized crime threat:** Cloud providers will store a range of different data types, including credit card and other financial and personal data. All of this data may be aggregated from multiple customers and therefore be extremely valuable to criminals. There is a risk that insiders are deliberately used to gain access to customer data and probe systems in order to assist any external attackers that require additional information in order to execute complex Internet-based attacks. Cloud customers should ensure that service providers are aware of this threat and have rigorous identity validation and security vetting procedures built into their recruitment process.

**Cheap data and data analysis:** The advent of cloud computing has created enormous data sets that can be monetized by applications such as advertising. Google, for example, leverages its cloud infrastructure to collect and analyze consumer data for its advertising network. Collection and analysis of data is now possible cheaply, even for companies lacking Google's resources. The availability of data and cheap data mining techniques have high impact on the privacy of user data. The attackers have massive, centralized databases available for analysis and also the raw computing power to mine these databases. Because of privacy concerns, enterprises running clouds for collecting data are increasingly finding the requirement of anonymizing their data. EPIC called for Gmail, Google Docs, Google Calendar, and the company's other web applications to shut down until appropriate privacy guards were in place (FTC Questions Cloud Computing Security). Google and Yahoo!, because of pressure from privacy advocates, now have an 18 months retention policy for their search data, after which they are to be anonymized by removing some identifiers of those data such as IP addresses and cookie information. The anonymized data is retained though, to support the continual testing of their algorithms. Another reason to anonymize data is to share with other parties for supporting research (AOL Incident, 2007) or to subcontract out data mining on the data (Netflix Prize). More sophisticated tools will be required for robust anonymization as increasing deployment of cloud applications takes place.

**Cost-effective defense of availability**: Availability also needs to be considered in the context of an adversary whose goals are simply to sabotage activities. Increasingly, such adversaries are becoming realistic as political conflict is taken onto the web, and as the recent cyber attacks on Lithuania confirm (Lithuania Weathers Cyber Attack). The damages are not only related to the losses of productivity but they degrade the trust in the infrastructure and make the backup processes more costly.

**Increased authentication demands:** The development of cloud computing may, in the extreme, allow the use of thin clients on the client side. Rather than purchasing a license and installing a software on the client side, users will authenticate in order to be able to use a cloud application. There are some advantages in such a model, such as making software piracy more difficult and making centralized monitoring more convenient. It may also help prevent the spread of sensitive data on untrustworthy clients. This architecture also supports enhanced mobility of users, but demands more robust authentication protocols. Moreover, the movement towards increased hosting of data and applications in the cloud and lesser reliance on specific user machines is likely to increase the threat of phishing and theft of access credentials.

**Mash-up authorization:** As adoption of cloud computing grows, more services performing mash-ups of data will be witnessed. This development has potential security implications, both in terms of data leaks, and in terms of the number of sources a data user may have to pull data from. This, in turn, places requirements on how access is authorized. A centralized access control mechanism may not be a feasible proposition in such deployment scenarios. One example in this area is provided by Facebook. Facebook users upload both sensitive and non-sensitive data. This data is used by Facebook to present the data to other users, and this data is also utilized by third party applications. Since these applications are typically not verified by Facebook, malicious applications running in Facebook's cloud can potentially steal sensitive data (Facebook Users Suffer Viral Surge, 2009).

## SOME PROPOSITIONS FOR SECURITY IN CLOUD COMPUTING

In this section, we discuss some novel security approaches that may be utilized in cloud computing deployments. The core issue is that with the advent of the cloud, the cloud provider also has some control of the cloud users' data. In this section, some propositions have been made in such a way that the current capabilities of the cloud are not curtailed while limiting the cloud provider control on data and enabling all cloud users to benefit from the cloud.

**Information-centric security:** In order for enterprises to extend control of data in the cloud, it may be worthwhile to take an approach of protecting data from within. This approach is known as information-centric security. This self-protection technique requires intelligence be put in the data itself. Data needs to be self-describing and defending, regardless of its environment. When accessed, data consults its policy and attempts to recreate a secure environment that is verified as trustworthy using the framework of *trusted computing* (TC).

**High-assurance remote server attestation:** At present, lack of transparency is discouraging businesses from moving their data to the cloud. Data owners wish to audit how their data is being handled at the cloud, and in particular, ensure that their data is not being abused or leaked, or at least have an unalterable audit trail when it does happen. Currently, customers must be satisfied with cloud providers using manual auditing procedures like SAS-70. A promising approach to address this problem is based on *trusted computing*. In a trusted computing environment, a trusted monitor is installed at the cloud server that can monitor or audit the operations of the cloud server. The trusted monitor can provide *proof of compliance* to the data owner, guaranteeing that certain access policies have not been violated. To ensure integrity of the monitor, trusted computing also allows secure bootstrapping of this monitor to run beside (and securely isolated from) the operating system and applications. The monitor can enforce access control policies and perform monitoring/auditing tasks. To produce a proof of compliance, the code of the monitor is signed, as well as a *statement of compliance* produced by the monitor. When the data owner receives this proof of compliance, it can verify that the correct monitor code is run, and that the cloud server has complied with access control policies.

**Privacy-enhanced business intelligence:** A different approach for retaining control of data is to require the encryption of all cloud data. The problem in this approach is that encryption limits data use. In particular, searching and indexing the data becomes problematic, if not impossible. For example, if data is stored in clear-text form, one can efficiently search for a document by specifying a keyword. This is impossible to do with traditional, randomized encryption schemes. The state-of-the-art cryptographic mechanisms may offer new tools to solve these problems. Cryptographers have invented versatile encryption schemes that allow for operations and computations on the cipher-texts. For example, *searchable encryption* (also referred to as *predicate encryption*) (Song et al., 2000) allows the data owner to compute a capability from his secret key. A capability encodes a search query, and the cloud can use this capability to decide which documents match the search query. The cloud can use this capability to decide which documents match the search query, without learning any additional information. Other cryptographic primitives such as *homomorphic encryption* (Gentry, 2009) and *private information retrieval* (PIR) (Chor et al., 1998) perform computations on encrypted source data without decrypting them. As these cryptographic techniques mature, they may open up new possibilities and directions for research, development and deployment of cloud security protocols and algorithms.

While in many cases more research is needed to make these cryptographic tools sufficiently practical for the cloud, they present the best opportunity for a clear differentiator for cloud computing since these mechanisms can enable cloud users to benefit from one another's data in a controlled manner. In particular, even encrypted data can enable anomaly detection that is valuable from a business intelligence standpoint. Apart from ensuring privacy, applied cryptography also offers tools to address other security problems related to cloud computing. For example, in *proofs of retrievability* (Shacham & Waters, 2008), the storage server can show a compact proof that it is correctly storing all of the client's data.

Table 5 summarizes some important security issues in cloud computing and their possible defense mechanisms.

Table 5. Cloud computing threats and suggested defense mechanisms for these threats

| Security threats | Possible defense mechanisms |
|---|---|
| Spoofing identity | Authentication |
| | Protect secrets |
| | Don't store secrets |
| Tampering with data | Authorization |

| | Hashes |
| | Message authentication codes |
| | Digital signatures |
| | Tamper-resistant protocols |
| Repudiation | Digital signatures |
| | Time-stamps |
| | Audit trails |
| Information disclosure | Authorization |
| | Privacy-enhanced protocols |
| | Encryption |
| | Protect secrets |
| | Don't store secrets |
| Denial of Service (DoS) | Authentication |
| | Authorization |
| | Filtering |
| | Throttling |
| | Quality of service (QoS) |
| Elevation of privilege | Run with least privilege |

## STANDARDIZATION ACTIVITIES IN CLOUD COMPUTING

This section presents a discussion on the various activities being undertaken by different standard development organizations (SDOs) in the world in the domain of cloud application and service deployments particularly with regards to security and privacy issues. For each standard development organization (SDO), we identify the focus on cloud computing related works particularly with regard to security and privacy issues.

### NIST Cloud Standards

NIST is a key organization in defining various standards for cloud computing. With regard to security and privacy aspects of cloud computing NIST has released standard guidelines for public clouds (Badger et al., 2011). The primary focus of the report issued by NIST is to provide an overview of public cloud computing and the security and privacy considerations involved. It discusses the threats, technology risks, and safeguards surrounding public cloud environments, and their suitable defense mechanisms. The report observes that "since the cloud computing has grown out of an amalgamation of technologies, including *service oriented architecture* (SOA), virtualization, Web2.0, and utility computing, many of the security and privacy issues involved in cloud computing can be viewed as known problems cast in a new setting" (Badger et al., 2011). However, public cloud computing manifests itself as a thought-provoking paradigm shift from conventional computing to an open deperimeterized organizational infrastructure – at the extreme, displacing applications form one organization's infrastructure to the infrastructure of another organization, where the applications of potential adversaries may also operate. The security and privacy issues which are identified by NIST to be relevant in cloud computing are: (i) governance, (ii) compliance, (iii) trust, (iv) hardware and software architecture, (v) identity and access management, (vi) software isolation, (vii) data protection, (viii) availability, and (ix) incident response.

Governance implies control and oversight by the organization over policies, procedures, and standards for application development and information technology service acquisition, as well as the design, implementation, testing, use, and monitoring of deployed or engaged services.

Compliance refers to an organization's responsibility to operate in agreement with established laws, regulations, standards, and specifications. Various types of security and privacy laws and regulation exist within different countries at the national, state, and local levels, making compliance a potentially complicated issue for cloud computing.

Trust is a critical issue in cloud computing since an organization relinquishes direct control over many aspects of security and privacy, and in doing so, confers a high level of trust onto the cloud provider. At the same time, federal agencies have a responsibility to protect information and information systems commensurate with the risk and magnitude of the harm resulting from unauthorized access, use, disclosure, disruption, modification, or destruction, regardless of whether the information is collected or maintained by or on behalf of the agency. In addition, the organization's ownership rights over the data must be firmly established in the service contract to enable a basis for trust and privacy of data.

The architecture of the software and hardware used to deliver cloud services can vary significantly among public cloud providers for any specific service model. Many cloud-based applications require a client side to initiate and obtain services. However, many of the simplified interfaces and service abstractions on the client, server, and network belie the inherent underlying complexity that affects security and privacy. Therefore, the NIST report recommends that it is important to understand the technologies the cloud provider uses to provision services and the implications the technical control involved have on security and privacy of the system throughout its lifecycle. With such information, the underlying system architecture of a cloud can be decomposed and mapped to a framework of security and privacy controls that can be used to access and manage risk. The hypervisor or virtual machine monitor is an additional layer of software between an operating system and hardware platform that is used to operate multi-tenant virtual machines and is common to IaaS clouds. Compared with traditional, non-virtualized implementation, the addition of hypervisor cause an increase in the attack surface in cloud computing, i.e., there are additional methods (e.g. application programming interfaces), channels (e.g., sockets), and data items (e.g., input strings) an attacker can use to cause damage to the system. The report of NIST recommends that care should be taken to provision security for the virtualized environments in which the images of various applications run. It also recommends the use of virtual firewalls to isolate groups of virtual machines from other hosted groups, such as production systems from development systems or development systems from other cloud-resident systems. Another aspect of security that is critical in cloud computing is the client-side protection. Since the services from different cloud providers, as well as cloud-based applications developed by the organization, can impose stringent demands on the client-side, which may have implications for security and privacy that need to be taken into account for system design. Likewise, the web browsers, which are key elements for many cloud computing services and various plug-ins and extensions available for them are notorious for their security problems. The growing availability and use of social media, personal webmail, and other publicly available sites also have associated risks that a concern, since they increasingly serve as avenues for social engineering attacks that can negatively impact the security of the browser, its underlying platform, and cloud services accessed. Since data sensitivity and privacy of information have become increasingly an area of concern for organizations in the paradigm of cloud computing, preventing unauthorized access to information resources in the cloud is a critical requirement. The NIST report recommends the use of identity federation as one solution to the complicated authentication requirements in cloud computing. Identity federation allows an organization and a cloud provider to trust and share digital identities and attributes across both domains, and to provide a means for single sign-on. For such federation to succeed, identity and access management transactions must be interpreted carefully and unambiguously and protected against attacks. There are several ways in which an identity federation can be accomplished such as with the *security assertion markup language* (SAML) standard or the OpenID standard (Badger et al., 2011). A growing number of cloud providers support the SAML standard and use it to administer users and authenticate them before providing access to application and data. SAML request and response messages are typically mapped over SOAP, which relies on the eXtensible Markup Language (XML) for its format. SOAP messages are digitally signed. In a public cloud, for instance, once a user has established a public key certificate with the service, the private key can be used to sign SOAP requests. However, SOAP message security validation is complicated and must be carried out carefully to prevent attacks. XML wrapping attacks have been successfully demonstrated against a public IaaS cloud (Gajek et al., 2009; Gruschka & Iacono, 2009). XML wrapping involves manipulation of SOAP messages. A new element (i.e., the wrapper) is introduces into the SOAP security header: the original message body is then moved under the wrapper and replaced by a bogus body containing an operation defined by the attacker (Gajek et al., 2009; Gruschka & Iacono, 2009). The original body can still be referenced and its signature verified, but the operation in the replacement body is executed instead. Since SAML alone is not sufficient to provide cloud-based identity and access management services, the NIST report recommends the use of eXtensible Access Control Markup Language (XACML) by a cloud provide to control access to cloud resources. XACML focuses on the mechanism for arriving at authorization decisions, which complements SAML's focus on the means for transferring authentication and authorization decisions between cooperating entities.

Since multi-tenancy in IaaS cloud computing environments is typically done by multiplexing the execution of virtual machines from potentially different consumers on the same physical server, applications deployed on guest virtual machines remain susceptible to attack and compromise, much the same as their non-virtualized counterparts (Badger et al., 2011). However, regardless of the service model and multi-tenant software architecture used, the computations of different consumers must be able to be carried out in isolation from one another, mainly through the use of logical separation mechanisms. This becomes an especially challenging proposition since multi-tenancy in virtual machine-based cloud infrastructures, together with the subtleties in the way physical resources are shared between guest virtual machine, can give rise to new sources of threat. The most serious threat is that malicious code can escape the confines of its virtual machine and interfere with the hypervisor or other guest virtual machines. Live migration, the ability to transition a virtual machine between hypervisors on different host computers without halting the guest operating system, and other features provided by virtual machine monitor environments to facilitate systems management, also increase software size and complexity and potentially add other areas to target in an attack.

Since the data stored in a public cloud typically resides in a shared environment collocated with data from other customers, the NIST report strongly recommends that access to the data should be controlled and the data should be kept secure (Badger et al., 2011). These requirements are also applicable for the data that is migrated within or between clouds. In addition, data can take many forms in the cloud. For example, for cloud-based application development, data may include the application programs, scripts, and configuration settings, along with the development tools. For developed applications, it includes records and other content created or used by the applications, including deallocated objects, as well as account information about the users of the applications.

The NIST report recommends two methods for keeping data away from unauthorized users: (i) access controls, and (ii) encryption. Access controls are typically identity-based, which makes authentication of the user's identity an important issue in cloud computing. However, lacking physical control over the storage of information, encryption is the only way to ensure that it is truly protected. In addition, data must be secured while at rest, in transit, and in use, and access to the data must be controlled. The standards for communication protocols and public key certificates allow data transfers to be protected using cryptography and can usually be implemented with equal effort in SaaS, PaaS, and IaaS environments (Badger et al., 2011). The NIST report observes that the security of a system that employs cryptography depends on the proper control of central keys and key management component. Currently, the responsibility for cryptographic key management falls mainly on the cloud consumer. Key generation and storage is usually performed outside the cloud using hardware security modules, which do not scale well to the cloud paradigm. NIST is currently undertaking the Cryptographic Key Management Project for identifying scalable and usable cryptographic key management and exchange strategies for use by government, which would help to alleviate the problem eventually (Cryptographic Key Management Project). NIST also recommends that before proceeding in cloud environments where the cloud provider provides facilities for key management, the organization must fully understand and weigh the risks involved in the processes defined by the cloud provider for the key management lifecycle (Badger et al., 2011). Hence, the cryptographic operations performed in the cloud become part of the key management process and, therefore should be managed and audited by the organization.

In a public cloud, data from one consumer is physically collocated (e.g., in an IaaS data store) or commingled (e.g., in a SaaS database) with the data of other consumers, which can complicate matters.

Hence, NIST recommends that sufficient measures should be taken to ensure that data sanitization should be performed appropriately throughout the system lifecycle.

The NIST report also observes that availability of services is a critical requirement for cloud service providers. Denial of service attacks, equipment outages, and natural disasters are all threats to availability. In most of these cases, the downtime is unplanned and can adversely affect the mission of the organization. Despite employing architectures designed for high service reliability and availability, cloud computing services can and do experience outages and performance slowdowns (Leavitt, 2009). NIST recommends that the level of availability of a cloud service and its capabilities for data backup and disaster recovery need to be addressed in the organization's contingency and continuity planning to ensure the recovery and restoration of disrupted cloud services and operations, using alternate services, equipment, and locations, if required.

NIST report points out the fact that a cloud service provider's role is vital in performing incident response activities, including incident verification, attack analysis, containment, data collection and preservation, problem remediation, and service restoration. Each layer in a cloud application stack, including the application, operating system, network, and database, generates event logs, as do other cloud components, such as load balancers and intrusion detection systems; many such event sources and the means of accessing them are under the control of the cloud service provider. The report also observes that availability of relevant data from event monitoring is essential for timely detection of security incidents. However, the cloud customers are often confronted with extremely limited capabilities for detection of incidents in public cloud environments and the service providers have insufficient access to event sources and vulnerability information, inadequate interfaces for accessing and processing event data automatically, and do not have the capability to add detection points within the cloud infrastructure, and have difficulty in directing third-party reported abuses and incidents effectively back to the correct customer or the cloud provider for handling. The report also observes that an incident should be handled in a way that limits damage and minimizes recovery time and costs. Hence, collaboration between the cloud consumer and provider in recognizing and responding to an incident is vital to security and privacy in cloud computing.

In summary, the NIST report on security and privacy issues in public cloud computing provides an overview of the public cloud and describes the threats, technology risks, and safeguards that are surrounding the public cloud environment. It also provides detailed guidelines for the service providers and the consumers to handle various security and privacy issues in cloud computing.

## Cloud Security Alliance (CSA)

This non-profit organization provides security guidance for critical areas of focus in cloud computing (CSA Homepage). The alliance covers key issues and provides advice for both cloud computing customers and providers within various strategic domains. CSA has published a report on cloud computing that outlines the areas of concern and guidance for organizations adopting cloud computing with an objective to provide the security practitioners with a comprehensive roadmap for being proactive in developing positive and secure relationships with cloud providers. The CSA guide on cloud computing deals with fifteen broad domains of cloud computing: (i) cloud computing architectural framework, (ii) governance and enterprise risk management, (iii) legal aspects of cloud computing, (iv) electronic discover, (v) compliance and audit, (vi) information lifecycle management, (vii) portability and interoperability issues, (viii) traditional security, business continuity and disaster recovery, (ix) data center operations, (x) incident response, notification and remediation, (xi) application security, (xii) encryption and key management, (xiii) identity and access management, (xiv) storage, and (xv) virtualization.

From security perspective, the CSA report recommends that a portion of the cost savings obtained by cloud computing services must be invested into the increased scrutiny of the security capabilities of the provider and ongoing detailed audits to ensure requirements are continuously met. It also recommends the following: (i) the service providers should have regular third party risk assessment and these should be made available to the customers, Iii) the cloud provider's key risk and performance indicators must be understood clearly and methods must be designed to monitor and measure these indicators from the perspective of the customers, (iii) the onus should be on the customer to perform due diligence of a cloud provider for usage in mission critical business functions or hosting regulated personally identifiable information, (iv) the cloud providers should adopt as a security baseline the most stringent requirements of any customer, (v) centralization of data implies the risk of insider threats from within the cloud provider is a significant concern, (vi) any data classified as private for the purpose of data breach regulations should always be encrypted to reduce the consequences of a breach incident and the customer should stipulate encryption requirements (algorithm, key length and key management at a minimum) contractually, (vii) IaaS, PaaS and SaaS create differing trust boundaries for the software development lifecycle, which must be accounted for during the development, testing and production deployment of applications, (viii) securing inter-host communications must be the rule, there can be no assumption of a secure channel between hosts, whether existing in a common data center or even on the same hardware platform, (ix) application providers who are not controlling backend systems should assure that data is encrypted when being stored on the backend, (x) segregate the key management from the cloud provider hosting the data, creating a chain of separation. This protects both the cloud provider and customer from conflict when being compelled to provide data due to a legal mandate.

## Distributed Management Task Force (DMTF)

DMTF develops standards for interoperable IT management solutions (DMTF Homepage). From this perspective DMTF is working on several topics like (1) open virtualization format (OVF) that formats for packaging and distributing software to run over virtual machines (2) Open Cloud Standards Incubator, for interactions between cloud environments by developing cloud resource management protocols. The activity was moved to Cloud Management Working Group (CMWG) and (3) Cloud Audit Data Federation (CADF) working group that develops solutions that allows sharing of audit information / logs.

For security issues in cloud computing, DMTF have established a partnership with CSA to promote standards for cloud security as part of DMTF Open Cloud Standard Incubator. The Open Cloud Standard Incubator group is charged with first formulating a series of management protocols, packaging formats and security tools to foster interoperability between cloud, followed by specifications that will foster cloud service portability and cross-cloud management consistency.

## Storage Networking Industry Association (SNIA)

SNIA has created the Cloud Storage Technical Work Group for the purpose of developing SNIA architecture related to system implementations of cloud storage technology (SNIA Homepage). It is promoting cloud storage as a new delivery model that provides elastic, on-demand storage billed only for what is used. The initiative, known as the Cloud Data Management Interface (CDMI), lets the customer to tag his/her data with special metadata (data system metadata) that the cloud storage provider what data services to provide that data (backup, archive, encryption etc.). These data services all add value to the data the customer stores in the cloud and by the implementation of the standard interface of CDMI, the customer can freely move his/her data from one cloud vendor to another without experiencing any pain of recoding to different interfaces.

SNIA is also involved in storage network security related activities. Although storage network security is a new subject, it is rapidly gaining in importance in the minds of both users and product developers. The increase is born of a general realization of the increasing importance and value of the information held in on-line systems, and of the separation of processing and storage functions enabled by the development of storage area networks (SANs). SINA's mission is "to ensure that storage networks become efficient, complete, and trusted solutions across the IT community". However, to achieve this goal, SNIA will have to develop new standards and technologies in storage network security. While storage network security seeks to learn from the application of similar techniques to communications security in general and to network security in particular, it has some unique requirements that will necessitate the development of new and specialized techniques. Currently, the development of such techniques is in its infancy.

## Open Grid Forum (OGF)

OGF's Open Cloud Computing Interface (OCCI) (OCCI Homepage) group creates practical solutions to interface with cloud infrastructures exposed as a service. The focus is on a solution which covers the provisioning, monitoring and definition of cloud infrastructure services.

The Open Cloud Computing Interface comprises a set of open community-led specifications delivered through the Open grid Forum. OCCI is a protocol and API for all kinds of management tasks. OCCI was originally initiated to create a remote management API for IaaS model based Services, allowing for the development of interoperable tools for common tasks including deployment, autonomic scaling and monitoring. It has since evolved into a flexible API with a strong focus on integration, portability, interoperability and innovation while still offering a high degree of extensibility. The current release of the Open Cloud Computing Interface is suitable to serve many other models in addition to IaaS, including e.g., PaaS and SaaS. The security group of OGF is concerned with technical and operational security issues in the grid and cloud environments, including authentication, authorization, privacy, confidentiality, auditing, firewalls, trust establishment, policy establishment, and dynamics, scalability and management aspects of these issues. The purpose of the Certificate Authority Operations (CAOPS) working group is to develop

## Open Cloud Consortium (OCC)

OCC (OCC Homepage) is a member driven organization that: (i) supports development of standards, (ii) supports development of benchmarks, (iii) supports reference implementations of cloud computing, preferably open source, and (iv) sponsors workshops and other events related to cloud computing. OCC has four working groups: (i) large data clouds working group, (ii) open cloud test-bed working group, (iii) standard cloud performance measurement (SCPM) working group, and (iii) information sharing and security working group. The SCPM working group is responsible for establishing benchmarks appropriate for four use cases: (i) moving an application between two clouds, (ii) obtaining burst instances from multiple cloud service providers for a private/public hybrid application, (iii) moving a large data cloud application to another large data cloud storage service, and (iv) moving a large data cloud application to another large data cloud computing service.

## Organization for the Advancement of Structured Information Standards (OASIS)

OASIS is a not-for-profit, international consortium that drives the development, convergence, and adoption of e-business standards (OASIS Homepage). The consortium produces more web services standards than any other organization along with standards for security, e-business, and standardization efforts in the public sector and application-specific markets. In cloud computing domain, OASIS has the following technical committees, each having its clearly defined objectives and goals.

1. Advanced Message Queuing Protocol (AMQP) TC: it defines a ubiquitous, secure, reliable and open ended Internet protocol for handling business messaging.
2. Cloud Application Management for Platforms (CAMP) TC: it is responsible for standardizing cloud PaaS management API.
3. Cloud Authorization (CloudAuthZ) TC: it focuses on enabling contextual attributes and entitlements to be delivered to Policy Enforcement Points in real-time.
4. Identity in the Cloud TC: it is involved in developing profiles of open standards for identity deployment, provisioning and management in cloud computing.
5. Open Data Protocol (OData) TC: its goal is to simplify data sharing across disparate applications in enterprise, cloud and mobile devices.
6. Privacy Management Reference Model (PMRM) TC: it is responsible for providing a guideline for developing operational solutions to privacy issues.
7. SOA Reference Model TC: it is involved in developing a core reference model to guide and foster the creation of specific service-oriented architecture (SOA).

8.  Topology and Orchestration Specification for Cloud Applications (TOSCA) TC: it is responsible for enhancing the portability of cloud applications and services.
9.  Transformational Government Framework TC: it is advancing an overall framework for using IT to improve delivery of public services.

## TM Forum

TM Forum (TM Forum Homepage) – an association that includes technology vendors such as HP and IBM, as well as more than 750 of the world's largest service providers in the communications, media and cloud service markets – has delivered what it calls the industry's first set of *enterprise-grade external compute Infrastructure as a Service (IaaS) requirements*. Put together by the association's Enterprise Cloud Leadership Council (ECLC), the document includes guidelines for technology; requirements for external private clouds in commercial, technical and operational terms; the business case for external private clouds; and sample use cases. It also details how business and technical agreements between enterprise customers and cloud service providers should be defined and managed to maximize benefits for both parties.

Focused on enabling best-in-class IT for service providers in the communications, media, defense and cloud service market, TM Forum created ECLC in 2009 to provide a forum for enterprise cloud users to share requirements and drive the development of best practices and standards that will remove the barriers to development and adoption of cloud services. Its list of members includes Deutsche Bank, Boeing, ING, Dassault Systems and Northrop-Grumman.

Incorporating the input from the top cloud innovators and thought leaders, this document of TM Forum intends to create a way forward for the industry that separates out the vital needs from the minor and secondary requirements. Based on end users' experience and requirements the document is intended to assist cloud service providers and technology suppliers to determine customer demands, drive direction on standards and best practices, and remove barriers to adoption. The vendors need to map their product and service offerings against those requirements.

The cloud services initiative of TM Forum, therefore, intends to deliver the following:

- An ecosystem of enterprise customers, cloud service providers and technology suppliers that enable the commercialization of this major business opportunity.
- Business guidance including benchmarks and service quality metrics.
- Technical agreements in collaboration with other industry groups.

The a particular focus on developing standards in cloud computing, the Enterprise Cloud Leadership Council (ECLC) of the TM Forum has the following programs in its agenda of activities:

- Defining service level agreements (SLAs) for cloud services
- Data-as-a-Service (DBaaS) reference architecture
- Cloud API requirements
- Business process and information frameworks for cloud
- Secure virtual private cloud reference architecture
- Standard service definitions/SKUs (Taxonomy of services)
- Cloud SDO liaisons
- eTOM and ITIL; how to combine them in a cloud context?
- Cloud service provider benchmarking and metrics
- Billing engine, client billing and partner revenue sharing for cloud services
- Common definition of commercial terms (business contract language)

The TM Forum has created a Cloud Services Initiative with the purpose to define a range of common approaches, processes, metrics and other key service enablers (TM Forum Homepage).

## International Telecommunication Union (ITU)

The International Telecommunications Union-Telecommunications Standards Group (ITU-T) (ITU-T Homepage) has formed a focus group on cloud computing (FG Cloud) to further ITU-T TSAG (Telecommunication Standardization Advisory Group) agreement at its meeting in Geneva during 8-11 February 2010. The focus group, established in accordance with Recommendation ITU-T A.7, from the standardization view points and within the competencies of ITU-T contributes to telecommunication aspects, i.e., the transport via telecommunications networks, security aspects of telecommunications, service requirements, etc., in order to support services/applications of cloud computing making use of telecommunication networks, specifically in the following activities:

- Identify of potential impacts on standard development and priorities for standards needed to promote and facilitate telecommunication/ICT support for cloud computing
- Investigate the need for future study items for fixed and mobile network in the scope of ITU-T
- Analyze which components would benefit most from interoperability and standardization
- Familiarize ITU-T and standardization communities with emerging attributes and challenges of telecommunication/ICT support for cloud computing
- Analyze the rate of change for cloud computing attributes, functions and features for the purpose of assessing the appropriate timing of standardization of telecommunication/ICT in support of cloud computing

The focus group on cloud computing in ITU-T collaborates with worldwide cloud computing communities (e.g., research institutions, laboratories, forums, and academia) including other SDOs and consortia. The group has also identified its specific tasks and deliverables in cloud computing standards development activities. The identified deliverables are: (i) identification of the benefits of cloud computing from telecommunication/ICT perspectives, (ii) gap analysis of ITU-T standards for telecommunication/ICT to support cloud computing, (iii) collection and summarization of vision and value propositions of cloud computing with a focus on telecommunication/ICT aspects, (iv) leveraging expertise within the ITU-T in building telecom networks to take advantage of cloud concepts and capabilities, (v) Analysis of telecommunication/ICT networking requirements functions and capabilities to support cloud computing services/applications (for both fixed and mobile devices), (vi) use case of services and reference models for telecommunication/ICT to support cloud computing, (vii) Designing the roadmap to guide further development of relevant ITU-T recommendations.

## The European Telecommunications Standards Institute (ETSI)

ETSI (ETSI Homepage) Technical Committee (TC) GRID, now known as TC CLOUD, has been formed to address issues associated with the convergence between IT (Information Technology) and Telecommunications. The focus is on scenarios where connectivity goes beyond the local network. This includes not only grid computing but also the emerging commercial trend towards cloud computing which places particular emphasis on ubiquitous network access to scalable computing and storage resources. Since TC CLOUD has particular interest in interoperable solutions in situations which involve contributions from both the IT and Telecom industries, the emphasis is on the Infrastructure as a Service (IaaS) delivery model.

The focus of the ETSI TC Cloud is on the following issues: (i) to complement progress being made elsewhere with a networking perspective and a more formal approach to standards and test specifications, (ii) introduce new requirements into networking (e.g., next-generation networks) standards to support new kinds of application such as grid and cloud, (iii) achieving the desired level of interoperability needed in next-generation networks, grids and clouds, (iv) collaborate with other SDOs in developing standards in cloud computing.

## Object Management Group (OMG)

OMG is an international, open membership, not-for-profit computer industry standards consortium. OMG Task Forces (TFs) develop enterprise integration standards for a wide range of industries. In cloud computing standardization, OMG's focus is on modeling deployment of applications and services on clouds for portability, interoperability and reuse (OMG Homepage). The standardization activities in cloud computing in OMG are mainly focused on the following broad areas:

- Meta-element association: for defining *distributed and non-deterministic computing* from the cloud and SOA perspective.
- Governance: there is a services governance domain and a cloud governance domain. The key is how to integrate these two points of view for governing *distributed and non-deterministic computing*.
- SLAs: developing SLAs for services delivered over the cloud
- SOA, events, and agents: defining communication among and within clouds between services enabled in these clouds.

## Association for Retail Technology Standards (ARTS)

ARTS (ARTS Homepage) is an international membership organization dedicated to reducing the costs of technology through standards. ARTS has been delivering application standards exclusively to the retail industry. ARTS released a white paper on cloud computing in 2009 that offers unbiased guidance for achieving maximum results from this relatively new technology. The version 1.0 of the whitepaper represents a significant update to the draft version released in October 2009. The document seeks to identify the characteristics of cloud computing that makes it compelling for retailers, and attempts to highlight areas in which a cloud-based solution offers strong benefits to retailers. It also discusses the key obstacles to adopting cloud-based solutions, including reliability, availability, and security. It also covers issues relating to portability, manageability, and interoperability.

## Institute of Electrical and Electronics Engineers (IEEE)

Hoping to propel cloud computing to new heights, the IEEE (IEEE Homepage) has launched a design guide and a standard for interoperable cloud services. According to IEEE, these two initiatives are by far the first ever attempt by any formal standards body to address the issues hounding cloud services. In order to enable transfer of customer data from one provider to another in a seamless standardized manner, the IEEE P2301 draft guide is being designed to provide an intuitive roadmap for application portability, management, and interoperability interfaces, as well as for file formats and operating conventions. The standard is expected to be completed in 2014 and will help vendors, service providers, and consumers involved in every aspects of procuring, developing, building, and using cloud computing.

In addition, IEEE is involved in preparation of another draft standard – IEEE P2302 draft standard for intercloud interoperability and federation. There is a growing demand from the consumers for the same kinds of global roaming, portability, and interoperability capabilities for storage and computing as with voice and text messaging. To meet this requirement, IEEE P2302 is defining the topology, protocols, functionality, and governance required for cloud-to-cloud interoperability.  The term "intercloud" refers to an interconnected mesh of clouds that depends on open standards for their operation. "Federation" allows users to move their data across internal and external clouds and access services running on other clouds according to the business and application requirements. The IEEE P2302 working group is also focusing on building a system among cloud product and service providers that would be transparent to users. The group plans to address transparent interoperability and federation in much the same way that standards do for the global telephony systems and the Internet.

**Alliance for Telecommunications Industry Solutions (ATIS)**

ATIS (ATIS Homepage) is the leading technical planning and standards development organization committed to the rapid development of global, market-driven standards for the information, entertainment and communications industry. ATIS' Cloud Services Forum (CSF) facilitates the adoption and advancement of cloud services from a network and IT perspective. Its primary focus is on the basic APIs in the control plane layer of the network rather than on the services on the network. Drawing upon business use cases that leverage cloud services' potential, CSF addresses industry priorities and develops implementable solutions for this evolving marketplace. CSF is working to ensure that services are quickly put into operation to facilitate the delivery of interoperable, secure and managed services. Current priorities of CSF include content distribution network interconnection, cloud services framework, inter-carrier tele-presence, virtual desktop, virtual private network, and development of a cloud services checklist for onboarding. The current initiatives of CSF include the following activities:

- Develop video service specifications as a component of a unified communications framework (e.g., telepresence, mobility etc.).
- Advance a trusted information exchange (TIE) solution to address the directory, routing, privacy, and accessibility.
- Progress the next phase of content distribution network-interconnection (CDN-I) – building on initial use cases to address more complicated models and additional content types. CSF currently leads the market in standardization aspects of CDN-I for content delivery, for example Multicast.
- Define virtual desktop functional requirements to take advantage of cloud resources to reduce management costs and support ay-device, any-network access to desktops by end-users.

**Internet Engineering Task Force (IETF)**

IETF (IETF Homepage) has established the Cloud OPS WG (working group on cloud computing and maintenance) which is currently discussing cloud resource management and monitoring, and Cloud-APS-BOF which has focused on cloud applications. There are several existing working groups within the IETF that are also working in the technical areas that could be useful to cloud computing activities. Among these working groups are Decade working group within IETF application area, nfsv4 working group within TSV application area, and netconf working group within OPS application area. Similarly, IRTF (Internet Research Task Force) has been working on the technical issues related to cloud computing as part of P2PRG working group and VNRG research group. While the above working groups have been in existence for some time, in the last year, there has been some renewed effort to focus on providing cloud services. Currently, there is an effort underway in the form of *birds of feather* (BOF) to discuss various contributions related to cloud computing. Most of the work being discussed as part of this effort would hopefully be very useful to the service providers.

## EMERGING TRENDS IN SECURITY AND PRIVACY IN CLOUD COMPUTING

Cloud computing environments are multidomain environments in which each domain can use different security, privacy, and trust requirements and potentially employ various mechanisms, interfaces, and semantics. Such domain could represent individually enabled services or other infrastructural or application components. Service-oriented architectures are naturally relevant technology to facilitate such multidomain formation through service composition and orchestration. It is important to leverage existing research on multidomain policy integration and the secure service composition to build a comprehensive policy-based management framework in cloud computing environments (Takabi et al., 2010). In the following, we identify some critical security and privacy issues in cloud computing that need immediate attention for ubiquitous adoption of this technology.

**Authentication and identity management:** By using cloud services, user can easily access their personal information and make it available to various services across the Internet. An identity management (IDM) mechanism can help authenticate users and services based on credentials and characteristics (Bertino et al., 2009). A key issue concerning IDM in cloud is interoperability drawbacks that could result from using different identity tokens and identity negotiation protocols. Existing password-based authentication has an inherited limitation and poses significant risks. An IDM system should be able to protect private and sensitive information related to users and processes. However, multi-tenant cloud environments can affect the privacy of identity information and isn't yet well understood. In addition, the multi-jurisdiction issue can complicate protection measures (Bruening & Treacy, 2009). While users interact with a front-end service, this service might need to ensure that their identity is protected from other services with which it interacts (Bertino et al., 2009; Ko et al., 2009). In multi-tenant cloud environments, providers must segregate customer identity and authentication information. Authentication and IDM components should also be easily integrated with other security components. Design and development of robust authentication and identity management protocols is a critical requirement for cloud computing.

**Access control and accounting:** Heterogeneity and diversity of services, as well as the domains' diverse access requirements in cloud computing environments, demand fine-grained access control policies. In particular, access control services should be flexible enough to capture dynamic, context, or attribute-or credential-based access requirements and to enforce the principle of least privilege. Such access control services might need to integrate privacy-protection requirements expressed through complex rules. It's important that access control system employed in clouds is easily managed and its privilege distribution is administered efficiently. It should also be ensured that the cloud delivery models provide generic access control interfaces for proper interoperability, which demands a policy-neutral access control specification and enforcement framework that can be used to address cross-domain access issues (Joshi et al., 2004). Utilizing a privacy-aware framework for access control and accounting services that is easily amenable to compliance checking is therefore a crucial requirement which needs immediate attention from the researches.

**Trust management and policy integration:** Although multiple service providers coexist in cloud and collaborate to provide various services, they might have different security approaches and privacy mechanisms. Hence, we must address heterogeneity among their policies (ENISA, 2009; Blaze et al., 2009; Zhang & Joshi, 2009). Cloud service providers might need to compose multiple services to enable bigger application services. Therefore, mechanisms are necessary to ensure that such a dynamic collaboration is handled securely and that security breaches are effectively monitored during the interoperation process. Existing work has shown that even though individual domain policies are verified, security violations can easily occur during integration (Zhang & Joshi, 2009). Hence, providers should carefully manage access control policies to ensure that policy integration doesn't lead to any security breaches. In cloud computing, the interactions between different service domains driven by service requirements can be dynamic, transient, and intensive. Thus, a trust framework should be developed to allow for efficiently capturing a generic set of parameters required for establishing trust and to manage evolving trust and interaction/sharing requirements (Zhang & Joshi, 2009; Shin & Ahn, 2005). In addition cloud's policy integration tasks should be able to address challenges such as semantic heterogeneity, secure interoperability, and policy evolution management. Since the customers' behavior can evolve rapidly, there is a need for integrated, trust-based, secure interoperation framework that helps establish, negotiate, and maintain trust to adaptively support policy integration. Design of efficient trust management frameworks for wireless and peer-to-peer networks is a widely researched problem (Sen, 2006c; Sen et al., 2007; Sen, 2010d; Sen, 2010e; Sen, 2010f; Sen, 2010g; Sen, 2011c). However, there is an urgent need for developing robust and reliable trust models for cloud computing environments. This will be a particularly challenging issue to address due to various interoperability issues and global deployments of cloud service delivery models.

**Secure service management:** In cloud computing environments, cloud service providers and service integrators compose services for their customers. The service integrator provides a platform that lets independent service providers orchestrate and interwork services and cooperatively provide additional services that meet customers' protection requirements. Although many cloud service providers use the Web Services Description Language (WSDL), the traditional WSDL can't fully meet the requirements of cloud computing services description. In clouds, issues such as quality of service, price, and SLAs are critical in service search and composition. These issues must be addressed to describe services and introduce their features, find the best interoperable options, integrate them without violating the service owner's policies, and ensure that SLAs are satisfied (Takabi et al., 2010). In essence, an automatic and systematic service provisioning and composition framework that considers security and privacy issues is crucial and needs urgent attention.

**Privacy and data protection:** Privacy is a core issue in many challenges in cloud computing including the need to protect identity information, policy components during integration, and transaction histories. Many organizations are not comfortable in storing their data and applications on systems that reside outside their on-premise data centers (Chen et al., 2010). By migrating workloads to a shared infrastructure, customers' private information faces increased risk of potential unauthorized access and exposure. Cloud service providers must assure their customers and provide a high degree of transparency into their operations and privacy assurance. Privacy-protection mechanisms must be embedded in all cloud security solutions. In a related issue, it's becoming important to know who created a piece of data, who modified it and how, and so on. Provenance information could be used for various purposes such as traceback, auditing, and history-based access control. Balancing between data provenance and privacy is a significant challenge in clouds where physical perimeters are abandoned. This is also a critical research challenge.

**Organizational security management:** Existing security management and information security life-cycle models significantly change when enterprises adopt cloud computing. In particular, shared governance can become a significant issue if not addressed properly. Despite the potential benefits of using clouds, it might mean less coordination among different communities of interest within client organizations. Dependence on external entities can also raise fears about timely response to security incidents and implementing systematic business continuity and disaster recovery plans. Similarly, risk and cost-benefit issues will need to involve external parties. Customers consequently need to consider newer risks introduced by a perimeter-less environment, such as data leakage within multi-tenant clouds and resiliency issues such as their provider's economic instability and local disasters. Similarly, the possibility of an insider threat is significantly extended when outsourcing data and processes to clouds. Within multi-tenant environments, one tenant could be a highly targeted attack victim, which could significantly affect the other tenant. Existing life-cycle models, risk analysis and management processes, penetration testing, and service attestation must be reevaluated to ensure that clients can enjoy the potential benefits of clouds (Takabi et al., 2010).

The information security area has faced significant problems in establishing appropriate security metrics for consistent and realistic measurements that help risk assessment. We must reevaluate best practices and develop standards to ensure the deployment and adoption of secure clouds. These issues necessitate a well-structured cyber insurance industry, but the global nature of cloud computing makes this prospect extremely complex. As well as trends specific to the cloud, general IT industry trends will also drive the change in cloud computing services and approach to future services, architectures and innovations (CPNI Security Briefing, 2010). Some of these trends are as follows:

**Increasing use of mobile devices:** Laptop sales have overtaken desktops over the last few years and the trend will continue as an increasing range of mobile devices such as notebooks, PDAs and mobile phones incorporate many of the features found on a desktop-based PC only about ten years ago, including Internet access and custom application functionality.

**Hardware capability improvements:** The inevitable improvements in processor speed and increased memory capacities across IT infrastructure will mean that the cloud will be able to support more complex environments with improved performance capabilities as standard.

**Tackling complexity:** Despite the efforts of multiple technology vendors, this challenge of complexity remains unresolved. IT architectures continue to be difficult to implement, under-utilized and expensive to operate. The massive scale of cloud computing only strengthens the need for self-monitoring, self-healing and self-configuring IT systems comprising heterogeneous storage, servers, applications, networks and other system elements.

**Legislation and security:** As larger companies consider the cloud computing model, vendors and providers will respond, but within the terms set out by their potential customers. As there are still many issues with respect to data privacy and transfer of data across international borders, the cloud service providers need to continue to invest time and effort in order to meet the necessary laws required to operate within some of the business areas of their major customers.

## CONCLUSION

Today, cloud computing is being defined and talked about across the ICT industry under different contexts and with different definitions attached to it. The core point is that cloud computing means having a server firm that can host the services for users connected to it by the network. Technology has moved in this direction because of the advancement in computing, communication and networking technologies. Fast and reliable connectivity is a must for the existence of cloud computing.

Cloud computing is clearly one of the most enticing technology areas of the current times due, at least in part to its cost-efficiency and flexibility. However, despite the surge in activity and interest, there are significant, persistent concerns about cloud computing that are impeding the momentum and will eventually compromise the vision of cloud computing as a new IT procurement model. Despite the trumpeted business and technical advantages of cloud computing, many potential cloud users have yet to join the cloud, and those major corporations that are cloud users are for the most part putting only their less sensitive data in the cloud. Lack of control is transparency in the cloud implementation – somewhat contrary to the original promise of cloud computing in which cloud implementation is not relevant. Transparency is needed for regulatory reasons and to ease concern over the potential for data breaches. Because of today's perceived lack of control, larger companies are testing the waters with smaller projects and less sensitive data. In short, the potential of the cloud is not yet being realized.

When thinking about solutions to cloud computing's adoption problem, it is important to realize that many of the issues are essentially old problems in a new setting, although they may be more acute (Chow et al., 2009). For example, corporate partnerships and offshore outsourcing involve similar trust and regulatory issues. Similarly, open source software enables IT department to quickly build and deploy applications, but at the cost of control and governance. Similarly, virtual machine attacks and web service vulnerabilities existed long before cloud computing became fashionable. Indeed, this very overlap is reason for optimism; many of these cloud computing roadblocks have long been studied and the foundations for solutions exist. For the enhancement of technology, and hence healthy growth of global economy, it is extremely important to iron out any issues that can cause road-blocks in this new paradigm of computing.

# REFERENCES

Alliance for Telecommunications Industry Solutions. Homepage URL: http://www.atis.org.

Amazon S3 Availability Event: (2008). URL: http://status.aws.amazon.com/s3-20080720.html (Accessed on November 29, 2012).

AOL Apologizes for Release of User Search Data (2006). URL: news.cnet.com/2010-1030_3-6102793.html. August 7, 2006.

Armbrust, M., Fox, A., Griffith, R., Joseph, A. D., Katz, R. H., Konwinsky, A., Lee, G., Patterson, D., Rabkin, A., Stoica, I., & Zaharia, M (2009). Above the Clouds: A Berkley View of Cloud Computing. Technical Report No. UCB/EECS-2009-28, Department of Electrical Engineering and Computer Sciences, University of California at Berkley. February 10, 2009. Available on line at: http://www.eecs.berkeley.edu/Pubs/TechRpts/2009/EECS-2009-28.pdf (Accessed on: November 20, 2012)

Association for Retail Technology Standards (ARTS). Homepage URL: http://www.nrf-arts.org.

Badger, L., Grance, T., Patt-Corner, R., & Voas, J. (2011). Draft Cloud Computing Synopsis and Recommendations. National Institute of Standards and Technology (NIST) Special Publication 800-146. US Department of Commerce. May 2011. Available online at: http://csrc.nist.gov/publications/drafts/800-146/Draft-NIST-SP800-146.pdf (Accessed on: November 20, 2012).

Bertion, E., Paci, F., & Ferrini, R. (2009). Privacy-Preserving Digital Identity Management for Cloud Computing. IEEE Computer Society Data Engineering Bulletin, pp. 1-4, March 2009.

Biggs & Vidalis (2009). Cloud Computing: The Impact on Digital Forensic Investigations. In Proceedings of the 7th International Conference for Internet Technology and Secured Transactions (ICITST'09), London, UK, November, 2009, pp. 1-6,

Blaze, M., Kannan, S., Lee I., Sokolsky, O., Smith, J. M., Keromytis, A.D., & Lee, W. (2009). Dynamic Trust Management. *IEEE Computer*, Vol 42, No 2, pp. 44-52, 2009.

Bruening, P.J. & Treacy, B.C. (2009). Cloud Computing: Privacy, Security Challenges. Bureau of National Affairs, 2009.

Center for the Protection of Natural Infrastructure (CPNI)'s Information Security Briefing on Cloud Computing, 01/2010, March 2010. Available Online at: http://www.cpni.gov.uk/Documents/Publications/2010/2010007-ISB_cloud_computing.pdf (Accessed on: November 29, 2012).

Chen, Y., Paxson, V., & Katz, R.H. (2010). What's New About Cloud Computing Security? Technical Report UCB/EECS-2010-5, EECS Department, University of California, Berkeley, 2010. Available Online at: http://www.eecs.berkeley.edu/Pubs/TechRpts/2010/EECS-2010-5.html (Accessed on: November 29, 2012).

Chor, B., Kushilevitz, E., Goldreich, O., & Sudan, M. (1998). Private Information Retrieval. *Journal of ACM* (*JACM*), Vol 45, No 9, pp. 965-981, November 1998.

Chow, R., Golle, P., Jakobsson, M., Shi, E., Staddon, J., Masuoka, R., & Molina, J. (2009). Controlling Data in the Cloud: Outsourcing Computation without Outsourcing Control. In *Proceedings of the ACM Workshop on Cloud Computing Security (CCSW'09)*, Chicago, Illinois, USA, November, 2009, pp 85-90, ACM Press, New York, USA.

Cloud Security Alliance. Home page URL: https://cloudsecurityalliance.org.

Cloud Security Alliance (CSA)'s Security Guidance for Critical Areas of Focus in Cloud Computing (2009). CSA, April 2009. Available Online at: https://cloudsecurityalliance.org/csaguide.pdf (Accessed on: November 29, 2012).

Cryptographic Key Management Project Website: URL: http://csrc.nist.gov/groups/ST/key_mgmt/ (Accessed on: November 29, 2012).

Distributed Management Task Force. Homepage URL: http://www.dmtf.org

Don't Cloud Your Vision. URL: http://www.ft.com/cms/s/0/303680a6-bf51-11dd-ae63-0000779fd18c.html?nclick_check=1. (Accessed on: November 29, 2012)

European Network and Information Security Agency (ENISA) (2009). Cloud Computing: Cloud Computing: Benefits, Risks and recommendations for Information Security. Report No: 2009.

European Telecommunication Standards Institute. Homepage URL: http://www.etsi.org.

Extended Gmail Outage Hits Apps Admins. (2008). URL: http://www.computerworld.com/s/article/9117322/Extended_Gmail_outage_hits_Apps_admins. October 16, 2008. (Accessed on: November 20, 2012)

Facebook Users Suffer Viral Surge. (2009). URL: http://news.bbc.co.uk/2/hi/technology/7918839.stm. March 2, 2009. (Accessed on: November 20, 2012)

Flexiscale Suffers 18-Hour Outage. (2008). URL: http://www.thewhir.com/web-hosting-news/flexiscale-suffers-18-hour-outage. October, 2008. (Accessed on: November 20, 2012).

FTC Questions Cloud Computing Security (2009). URL: http://news.cnet.com/8301-13578_3-10198577-38.html?part=rss&subj=news&tag=2547-1_3-0-20. (Accessed on: November 29, 2012).

Gajek, S., Jensen, M., Liao, L., & Schwenk, J. (2009). Analysis of Signature Wrapping Attacks and Countermeasures. In *Proceedings of the IEEE International Conference on Web Services*, Los Angeles, California, USA, July 2009, pp. 575-582.

Garfinkel, S. & Shelat, A. (2003). Remembrance of Data Passed: A Study of Disk Sanitization Practices. *IEEE Security and Privacy*, Vol 1, No 1, pp. 17-27, January-February 2003.

Gartner Hype-Cycle 2012 – Cloud Computing and Big Data (2012). Available at: http://www.gartner.com/technology/research/hype-cycles/

Gentry, C. (2009). Fully Homomorphic Encryption Using Ideal Lattices. In *Proceedings of the 41st Annual ACM Symposium on Theory of Computing (STOC'09)*, pp. 169-178, Bethesda, Maryland, USA, May-June, 2009.

Gruschka, N. & Iacono, L. L. (2009). Vulnerable Cloud: SOAP Message Security Validation Revisited. In *Proceedings of IEEE International Conference on Web Services (ICWS'09)*, Los Angeles, California, USA, July 2009, pp. 625-631.

IBM Blue Cloud Initiative Advances Enterprise Cloud Computing. URL: http://www-03.ibm.com/press/us/en/pressrelease/26642.wss. (Accessed on: November 20, 2012).

Institute of Electrical and Electronics Engineers (IEEE). Homepage URL: http://www.ieee.org.

International Telecommunication Union – Telecommunication Standardization Sector (ITU-T). Homepage URL: http://www.itu.int/ITU-T.

Internet Engineering Task Force. Homepage URL: http://www.ietf.org

Joshi, J.B.D., Bhatti, R., Bertino, E., & Ghafoor, A. (2004). Access Control Language for Multi-domain Environments. *IEEE Internet Computing*, Vol 8, No 6, pp. 40-50, November 2004.

Ko, M., Ahn, G.-J., & Shehab, M. (2009). Privacy-Enhanced User-Centric Identity Management. In *Proceedings of IEEE International Conference on Communications*, Dresden, Germany, June 2009, pp. 998-1002.

Latest Cloud Storage Hiccups Prompts Data Security Questions. URL:http://www.computerworld.com/action/article.do?command=viewArticleBasic&articleId=9130682&source=NLT_PM. (Accessed on: November 29, 2012)

Leavitt, N. (2009). Is Cloud Computing Really Ready for Prime Time? *IEEE Computer*, Vol 42, No 1, pp. 15-20, January 2009.

Leighon, T. (2009). Akamai and Cloud Computing: A Perspective from the Edge of the Cloud. White Paper. Akamai Technologies. Available online at: http://www.essextec.com/assets/cloud/akamai/cloud-computing-perspective-wp.pdf. (Accessed on: November 20, 2012).

Lithuania Weathers Cyber Attack, Braces for Round 2. URL: http://blog.washingtonpost.com/securityfix/2008/07/lithuania_weathers_cyber_attac_1.html. (Accessed on: November 20, 2012).

Loss of Customer Data Spurs Closure of Online Storage Service The Linkup'. URL: http://www.networkworld.com/news/2008/081108-linkup-failure.html?page=1. (Accessed on: November 20, 2012).

Lowensohn, J. & McCarthy, C. (2009). Lessons from Twitter's Security Breach. Available online at: http://news.cnet.com/8301-17939_109-10287558-2.html (Accessed on: November 29, 2012).

Microsoft Security Bulletin MS07-049. Vulnerability in Virtual PC and Virtual Server Could Allow Elevation of Privilege (937986). URL: http://www.microsoft.com/technet/security/bulletin/ms07-049.mspx. (November 13, 2007) (Accessed on November 20, 2012).

Molnar, D. & Schechter, S. (2010). Self Hosting vs. Cloud Hosting: Accounting for the Security Impact of Hosting in the Cloud. In Proceedings of the Workshop on the Economics of Information Security, 2010, Harvard University, USA, June 2010. Available online at: http://weis2010.econinfosec.org/papers/session5/weis2010_schechter.pdf (Accessed on: November 29, 2012).

Netflix Prize. URL: http://www.netflixprize.com/

Object Management Group. Homepage URL: http://www.omg.org.

Open Cloud Computing Interface. Homepage URL: http://occi-wg.org.

Open Cloud Consortium. Homepage URL: http://opencloudconsortium.org.

Organization for the Advancement of Structured Information Standards. Homepage URL: http://www.oasis-open.org.

Ristenpart, T., Tromer, E., Shacham, H., & Savage, S. (2009). Hey, You, Get Off Of My Cloud: Exploring Information Leakage in Third-Party Compute Clouds. In *Proceedings of the 16th ACM Conference on Computer and Communications Security (CCS'09)*, November, 2009, Chicago, Illinois, USA, pp. 199-212. ACM Press, New York, USA, 2009.

Salesforce.com Warns Customers of Phishing Scam. (2007) URL: http://www.pcworld.com/businesscenter/article/139353/article.html. November, 2007 (Accessed on: November 20, 2012).

Security Evaluation of Grid Environments. Available online at: http://www.slideworld.com/slideshows.aspx/Security-Evaluation-of-Grid-Environments-ppt-217556. (Accessed on: November 29, 2012)

Security Tracker. VMWare Shared Folder Bug Lets Local Users on the Guest OS Gain Elevated Privileges on the Host OS. Security Tracker ID: 1019493. URL: http://securitytracker.com/id/1019493 (Accessed on: November 20, 2012)

Sen, J. (2011a). A Robust Mechanism for Defending Distributed Denial of Service Attacks on Web Servers. *International Journal of Network Security and its Applications*, Vol 3, No 2, pp. 162-179, March 2011.

Sen, J. (2011b). A Novel Mechanism for Detection of Distributed Denial of Service Attacks. In *Proceedings of the 1st International Conference on Computer Science and Information Technology (CCSIT'11)*, pp. 247-257, Springer CICS Vol 133, Part III, January 2011, Bangalore, India.

Sen, J. (2010a). An Agent-Based Intrusion Detection System for Local Area Networks. *International Journal of Communication Networks and Information Security (IJCNIS)*, Vol 2, No 2, pp. 128-140, August 2010.

Sen, J. (2010b). An Intrusion Detection Architecture for Clustered Wireless Ad Hoc Networks. In *Proceedings of the 2nd IEEE International Conference on Intelligence in Communication Systems and Networks (CICSyN'10)*, pp. 202-207, July, 2010, Liverpool, UK.

Sen, J. (2010c). A Robust and Fault-Tolerant Distributed Intrusion Detection System. In *Proceedings of the 1st International Conference on Parallel, Distributed and Grid Computing (PDGC'10)*, pp. 123-128, October 2010, Waknaghat, India.

Sen, J. (2010d). A Distributed Trust Management Framework for Detecting Malicious Packet Dropping Nodes in a Mobile Ad Hoc Network. *International Journal of Network Security and its Applications (IJNSA)*, Vol 2, N0 4, pp. 92-104, October 2010.

Sen, J. (2010e). A Distributed Trust and Reputation Framework for Mobile Ad Hoc Networks. In *Proceedings of the 1st International Workshop on Trust Management in Peer-to-Peer Systems (IWTMP2PS)*, pp. 538-547, July 2010, Chennai, India, Springer CCIS Vol 89.

Sen, J. (2010f). A Trust-Based Robust and Efficient Searching Scheme for Peer-to-Peer Networks. In *Proceedings of the 12th International Conference on Information and Communication Security (ICICS)*, pp. 77-91, December 2010, Barcelona, Spain, Springer LNCS Vol 6476.

Sen, J. (2010g). Reputation- and Trust-Based Systems for Wireless Self-Organizing Networks. Book Chapter in *Security of Self-Organizing Networks: MANET, WSN, WMN, VANET*, pp. 91-122, Al-Shakib Khan Pathan et al. (eds.), Aurbach Publications, CRC Press, USA, December 2010.

Sen, J. (2011c). A Secure and Efficient Searching for Trusted Nodes in Peer-to-Peer Network. In *Proceedings of the 4th International Conference on Computational Intelligence in Security for Information Systems (CISIS'11)*, pp. 101-109, Springer LNCS Vol 6694, June 2011.

Sen, J. & Sengupta, I. (2005). Autonomous Agent-Based Distributed Fault-Tolerant Intrusion Detection System. In *Proceedings of the 2nd International Conference on Distributed Computing and Internet Technology (ICDCIT'05)*, pp. 125-131, December, 2005, Bhubaneswar, India. Springer LNCS Vol 3186.

Sen, J., Sengupta, I., & Chowdhury, P. R. (2006a). A Mechanism for Detection and Prevention of Distributed Denial of Service Attacks. In *Proceedings of the 8th International Conference on Distributed Computing and Networking (ICDCN'06)*, pp. 139-144, Springer LNCS Vol 4308, December 2006, Guwahati, India.

Sen, J., Sengupta, I., & Chowdhury, P.R. (2006b). An Architecture of a Distributed Intrusion Detection System Using Cooperating Agents. In *Proceedings of the International Conference on Computing and Informatics (ICOCI'06)*, pp. 1-6, June, 2006, Kuala Lumpur, Malaysia.

Sen, J., Chowdhury, P. R., & Sengupta, I. (2006c). A Distributed Trust Mechanism for Mobile Ad Hoc Networks. In *Proceedings of the International Symposium on Ad Hoc and Ubiquitous Computing (ISAHUC'06)*, pp. 62-67, December, 2006, Surathkal, Mangalore, India.

Sen, J., Chowdhury, P. R., & Sengupta, I. (2007). A Distributed Trust Establishment Scheme for Mobile Ad Hoc Networks. In *Proceedings of the International Conference on Computation: Theory and Applications (ICCTA'07)*, pp. 51-57, March 2007, Kolkata, India.

Sen, J., Ukil, A., Bera, D., & Pal, A. (2008). A Distributed Intrusion Detection System for Wireless Ad Hoc Networks. In *Proceedings of the 16th IEEE International Conference on Networking (ICON'08)*, pp. 1-5, December 2005, New Delhi, India.

Sinclair, S. & Smith, S. W. (2008). Preventive Directions for Insider Threat Mitigation Using Access Control. Book Chapter No 11, S. Stolfo, S. M. Bellovin, S. Hershkop, A. D. Keromytis, S. Sinclair, & W. Smith eds. *Insider Attack and Cyber Security: Beyond the Hacker*. Springer, April 2008.

Shacham, H. & Waters, B.  (2008). Compact Proofs of Retrievability. In *Proceedings of the 14th International Conference on the Theory and Application of Cryptology and Information Security: (ASIACRYPT'08)*, Melbourne, Australia, December 7-11, 2008. Lecture Notes in Computer Science (LNCS), Vol 5350, pp. 90-107, Springer-Verlag, Berlin, Heidelberg, Germany, 2008.

Shin, D. & Ahn, G.-J. (2005). Role-Based Privilege and Trust Management. *Computer Systems Science and Engineering Journal*, Vol 20, No 6, pp. 401-410, November 2005.

Song, D., Wagner, D., & Perrig, A. (2000). Practical Techniques for Searches on Encrypted Data. In *Proceedings of the IEEE Symposium on Research in Security and Privacy*, Oakland, California, USA, pp. 44-55, May 2000.

Storage Networking Industry Association. Homepage URL: http://www.snia.org.

Takabi, H., Joshi, J. B. D., & Ahn, G.-J. (2010). Security and Privacy Challenges in Cloud Computing Environments. *IEEE Security and Privacy*, Vol 8, No 6, pp. 24-31, November-December 2010.

TM Forum. Homepage URL: http://www.tmforum.org.

Trusted Computing Group (TCG)'s White Paper (2010). Cloud Computing and Security- A Natural Match.  Available online at: http://www.trustedcomputinggroup.org (Accessed on; November 2012).

Xen Vulnerability. URL: http://secunia.com/advisories/26986/. (Accessed on: November 20, 2012).

Zetter, K. (2010). Google hackers Targeted Source Code of More Than 30 Companies. Wired Threat Level. January 13 2010. Available online at: http://www.wired.com/threatlevel/2010/01/google-hack-attack/ (Accessed on: November 29, 2012).

Zhang, Y. & Joshi, J. (2009). Access Control and Trust Management for Emerging Multidomain Environments. *Annals of Emerging Research in Information Assurance, Security and Privacy Services*, S. Upadhyay and R.O. Rao (eds.), Emerald Group Publishing, pp. 421-452, 2009.

## KEY TERMS AND DEFINITIONS

**Cloud Computing:** As per the definition provided by the National Institute for Standards and Technology (NIST), USA, Cloud Computing is defined as "a model for enabling convenient, on-demand network access to a shared pool of configurable computing resources (e.g., networks, servers, storage, applications, and services) that can be rapidly provisioned and released with minimal management effort or service provider interaction".

**Public Cloud:** The Public Clouds are provided by a designated service provider and may offer either a single-tenant (dedicated) or multi-tenant (shared) operating environment with all the benefits and functionality of elasticity and the accountability/utility model of cloud. The physical infrastructure is generally owned by and managed by the designated service provider and located within the provider's data centers (off-premises). All customers share the same infrastructure pool with limited configuration, security protections, and availability variances.

**Private Cloud:** The Private Clouds are provided by an organization or its designated services that offers a single-tenant (dedicated) operating environment with all the benefits and functionality of elasticity and accountability/utility model of cloud. The private clouds aim to address concerns on data security and offer greater control, which is typically lacking in a public cloud.

**Hybrid Cloud:** The Hybrid Clouds are a combination of public and private cloud offerings that allow for transitive information exchange and possibly application compatibility and portability across disparate cloud service offerings. They utilize standard or proprietary methodologies regardless of the ownership or the location. With a hybrid cloud, the service providers can utilize third party cloud providers in a full or partial manner, thereby increasing the flexibility of computing.

**Software as a Service (SaaS):** In this cloud service delivery model, the capability provided to the consumer is to use the provider's applications running on a cloud infrastructure and accessible from various client devices through a thin client interface such as web browser. A complete application is offered to the customer as a service on demand. On the customers' side, there is no need for upfront investment in servers or software licenses, while for the provider, the costs are lowered, since only a single application needs to be hosted and maintained. Currently, SaaS is offered by companies such as Google, Salesforce, Microsoft, Zoho etc.

**Platform as a Service (PaaS):** In this cloud service model, a layer of software or development environment is encapsulated and offered as a service, upon which the higher levels of services are built. The customer has the freedom to build his own applications, which run on the provider's infrastructure. Hence, a capability is provided to the customer to deploy onto the cloud infrastructure customer-created applications using programming languages and tools supported by the provider (e.g., Java, Python, .Net etc.). Although the customer does not manage or control the underlying cloud infrastructure, network, servers, operating systems, or storage, but he/she has the control over the deployed applications and possibly over the application hosting environment configurations. Some examples of PaaS are: Google's App Engine, Force.com, etc.

**Infrastructure as a Service (Iaas):** This cloud service delivery model provides basic storage and computing capabilities as standardized services over the network. Servers, storage systems, networking equipment, data center space etc. are pooled and made available to handle workloads. The capability provided to the customer is to rent processing, storage, networks, and other fundamental computing resources where the customer is able to deploy and run arbitrary software, which can include operating systems and applications. The customer does not manage or control the underlying cloud infrastructure but has the control over operating systems, storage, deployed applications, and possibly select networking components (e.g., firewalls, load balancers etc.). Some examples of IaaS are: Amazon, GoGrid, 3 Tera etc.

**Virtual Machine (VM):** A virtual machine (VM) is a software implementation of a computing environment in which an operating system or a program can be installed and run. It typically emulates a physical computing environment, but requests for CPU time slot, memory,, hard disk, network and other resources are managed by a virtualization layer which translates these requests to the underlying physical hardware. VMs are created within a virtualization platform that runs on top of a client or server operating system. This operating system is known as the host operating system. The virtualization layer can be used to create many individual, isolated VM environments.

**Hypervisor:** A hypervisor, also called a virtual machine manager, is a program that allows multiple operating systems to share a single hardware host. Each operating system appears to have the host's processor, memory, and other resources all to itself. However, the hypervisor is actually controlling the host processor and resources, allocating what are needed to each operating system in turn and making sure that the guest operating systems (i.e., the virtual machines) cannot disrupt each other.

**Trusted Platform Module (TPM):** The Trusted Platform Module (TPM) is a component on a computing machine that is specifically designed to enhance platform security above-and-beyond the capabilities of security software by providing a protected space for key operations and other critical security-related tasks. Using both hardware and software modules, TPM protects encryption and signature keys at their most vulnerable stages-operations when the keys are being used unencrypted in plain-text form.

**Side Channel Attack:** A side channel attack is any attack based on information gained from the physical implementation of a system; e.g., timing information, power consumption, electromagnetic leaks or even sound can provide an extra source of information that can be exploited to access or damage the system. Since these attacks are non-invasive, passive and they can generally be performed using relatively cheap equipment, they pose a serious threat to the security of most cryptographic hardware devices.

**Homomorphic Encryption:** Homomorphic Encryption is a form of encryption that allows specific types of computations to be carried out on ciphertext and obtain an output that is the result of operations performed on the plaintext. For example, one person could add two encrypted numbers and then another person could decrypt the result, without either of them being able to find the value of the individual numbers. The homomorphic property of various cryptosystems can be used to create secure voting systems, collision-resistant hash functions, private information retrieval schemes and enable widespread use of cloud computing by ensuring the confidentiality of processed data.

**Security Assertion Markup Language (SAML):** It is a standard for exchanging authentication and authorization data between security domains. Essentially the standard is based on an XML (eXtended Markup Language)-based protocol that uses security tokens containing assertions to pass information about a principal (usually an end user) between an SAML authority, that is an identity provider, and a web service, that is a service provider. SAML 2.0 standard enables web-based authentication and authorization scenarios including single sign-on (SSO).

**Advanced Message Queuing Protocol (AMQP):** AMQP is an open standard application layer protocol for message-oriented middleware. The defining features of AMQP are message orientation, queuing, routing, reliability, and security. It defines a ubiquitous, secure, reliable and open ended Internet protocol for handling business messaging.

**Identity Management:** Identity Management is the task of controlling information about users on computing machines. Such information includes information that authenticates the identity of a user, information that describes information and actions they are authorized to access and/or perform. It also includes the management of descriptive information about the user and how and by whom the information can be accessed and modified. Managed entities typically include hardware and network resources and even applications.

**Web Services Description Language (WSDL):** WSDL is an XML format describing network services as a set of endpoints operations on messages containing either document-oriented or procedure-oriented information. The operations and messages are described abstractly and then bound to a concrete network protocol and message format to define an endpoint. Related endpoints are combined into abstract endpoints 9services). WSDL is extensible to allow description of endpoints and their messages regardless of what message formats or network protocols are used to communicate.

## ADDITIONAL READINGS

Barham, P., Dragovic, B., Fraser, K., Hand, S., Harris, T., Ho, A., Neugebauer, R., Pratt, I., and Warfield, A. (2003). Xen and the Art of Virtualization. *Technical Report*, University of Cambridge. Available online at: www.cl.cam.ac.uk/research/srg/netos/papers/2003-xensosp.pdf (Accessed on: January 23, 2013).

Bhattacherjee, B., Abe, N., Goldman, K., Zadrozny, B., Chillakuru, V. R., Del Caprio, M., and Apte, C. (2006). Using Secure Coprocessors for Privacy Preserving Collaborative Data Mining and Analysis. In *Proceedings of the 2ⁿᵈ International Workshop on Data Management on New Hardware (DaMoN'06)*, Chicago, Illinois, USA, June, 2006 Article No 1, New York: ACM Press.

Boneh, D., and Waters, B. (2007). Conjunctive, Subset, and Range Queries on Encrypted Data. In *Proceedings of the 4ᵗʰ Conference on Theory of Cryptography (TCC'07)*, pp. 53-534.

Brandic, I., Music, D., Leitner, P., Dustdar, S. (2009). VieSLAF Framework: Enabling Adaptive and Versatile SLA-Management. In *Proceedings of the 6ᵗʰ International Workshop on Grid Economics and Business Models (GECON'09)*, pp. 60-73, August 25-28, 2009, Delft, The Netherlands.

Cavoukian, A. (2008). Privacy in the Clouds: A White Paper on Privacy and Digital Identity: Implications for the Internet. Available online at: http://www.ipc.on.ca/images/resources/privacyintheclouds.pdf.

Chappel, D. (2008). Introducing the Azure Services Platform. Available online at: http://download.microsoft.com.(Accessed on: January 23, 2013).

Chong, F., Carraro, G., and Wolter, R. (2006). Multi-Tenant Data Architecture. Available online at: http://msdn.microsoft.com/en-us/library/aa479086.aspx. (Accessed on: January 23, 2013).

Cloud Computing Security: Making Virtual Machines Cloud Ready. Available online at: http://www.techrepublic.com/whitepapers/cloud-computing-security-making-virtual-machines-cloud-ready/1728295. (Accessed on: January 23, 2013).

Creeger, M. (2009). Cloud Computing: An Overview. *ACM Queue- Distributed Computing*, Vol 7, Issue 5, p. 2, June 2009. New York: ACM Press.

DeCandia, G., Hastorun, D., Jampani, M., Kakulapati, G., Lakshman, A., Pilchin, A., Sivasubramanian, S., Vosshall, P., and Vogels, W. (2007). Dynamo: Amazon's Highly Available Key-Value Store. In *Proceedings of the 21ˢᵗ ACM SIGOPS Symposium on Operating Systems Principles (SOSP'07)*, pp. 205-220, Stevenson, WA, USA, October 2007.

Desisto, R. P., Plummer, D. C., and Smith, D. M. (2008). Tutorial for Understanding the Relationship between Cloud Computing and SaaS. Stamford, CT: Gartner, April 2008.

Emig, C., Brandt, F., Kreuzer, S., and Abeck, S. (2007). Identity as a Service- Towards a Service-Oriented Identity Management Architecture. In *Proceedings of the 13ᵗʰ Open European Summer School and IFIP TC6.6 Conference on Dependable and Adaptable Network and Services (EUNICE'07)*, pp. 1-8, July 2007, Twente, The Netherlands.

Everett, C. (2009). Cloud Computing- A Question of Trust. *Computer Fraud & Security*, Vol 2009, Issue 6, pp. 5-7 June 2010.

Gellman, R. (2009). Privacy in the Clouds: Risks to Privacy and Confidentiality from Cloud Computing. World Privacy Forum (WPF) REPORT, February 23, 2009. Available online at: http://www.worldprivacyforum.org/cloudprivacy.html (Accessed on: January 23, 2013).

Golden, B. (2009). Capex vs. Opex: Most People Miss the Point about Cloud Economics. URL: http://www.cio.com/article/484429/Capex_vs._Opex_Most_People_Miss_the_point_About_Cloud_Economic.

Heritage, T. (2009). Hosted Informatics: Bringing Cloud Computing Down to Earth with Bottom-Line Benefits for Pharma. *Next Generation Pharmaceutical*, Issue 17, October 2009.

Itani, W., Kayssi, A., and Chehab, A. (2009). Privacy as a Service: Privacy-Aware Data Storage and Processing in Cloud Computing Architectures. In *Proceedings of the 8ᵗʰ IEEE International Conference on Dependable, Automatic and Secure Computing (DASC'09)*, pp. 711-716, Chengdu, China, December 2009.

Kaufman, L. M. (2009). Data Security in the World of Cloud Computing. *IEEE Security & Privacy*, Vol 7, Issue 4, pp. 61-64, July-August 2009.

Messmer, E. (2009). Gartner on Cloud Security: 'Our Nightmare Scenario is Here Now.' *Network World* October 21, 2009. URL: http://www.networkworld.com/news/2009/102109-gartner-cloud-security.html. (Accessed on: January 23, 2013).

Open Cloud Manifesto. http://www.opencloudmanifesto.org/Open%20Cloud%20Manifesto.pdf. (Accessed on: January 23, 2013).

Pearson, S. (2009). Taking Account of Privacy when Designing Cloud Computing Services. In *Proceedings of the ICSE Workshop on Software Engineering Challenges of Cloud Computing (CLOUD'09)*, pp. 44-52, Vancouver, British Columbia, Canada, May 2009.

Pearson, S. and Charlesworth, A. (2009). Accountability as a Way Forward for Privacy Protection in the Cloud. In *Proceedings of the 1ˢᵗ International Conference on Cloud Computing (CloudCom'09)*, pp. 131-144, December 2009, Beijing, China.

Petry, A. (2007). Design and Implementation of a Xen-Based Execution Environment. *Diploma Thesis,* Technische Universitat Kaiserslautern, April 2007.

Price, M. (2008). The Paradox of Security in Virtual Environments. *IEEE Computer*, Vol 41, Issue 11, pp. 22-38, November 2008.

RightScale Inc. (2009). RightScale Cloud Management Features. URL: http://www.rightscale.com/products/cloud-management.php. (Accessed on: January 23, 2013).

Rochwerger, R., Caceres, J., Montero, R. S., Breitgand, D., Elmroth, E., galls, A., Levy, E., Llorente, I. M., Nagin, K., and Wolfsthal, Y. (2009). The RESERVOIR Model and Architecture for Open Federated Cloud Computing. *IBM Systems Journal*, September 2009.

Schubert, L., Kipp, A., and Wesner, S. (2009). Above the Clouds: From Grids to Service-Oriented Operating Systems. In G. Tselentis, Jet al. (Eds.), *Towards the Future Internet- A European Research Perspective*, pp. 238-249, Amsterdam: IOS Press.

Sims, K. (2009). IBM Blue Cloud Initiative Advances Enterprise Cloud Computing. URL: http://www-03.ibm.com/press/us/en/pressrelease/26642.wss. (Accessed on: January 23, 2013).

Sotomayor, B., Montero, R. S., Llorente, I. M., and Foster, I. (2009). Virtual Infrastructure Management in Private and Hybrid Cloud. *IEEE Internet Computing*, Vol 13, Issue 5, pp. 14-22, September-October 2009.

Vaquero, L. M., Rodero-Merino, L., Caceres, J., and Linder, M. (2009). A Break in the Clouds: Towards a Cloud Definition. *ACM SIGCOMM Computer Communication Review*, Vol 39, Issue 1, pp. 50-55, January 2009.

Vouk, M. A. (2008). Cloud Computing – Issues, Research and Implementations. In *Proceedings of the 30ᵗʰ International Conference on Information Technology Interfaces (ITI'08)*, pp. 31-40, Cavtat, Croatia, June 2008.

Vozmediano, R. M., Montero, R. S., and Llorente, I. M. (2011). Multi-Cloud Deployment of Computing Clusters for Loosely-Coupled MTC Applications. *IEEE Transactions on Parallel and Distributed Systems*, Vol 22, Issue 6, pp. 924-930.

Zimory GmbH (2009). Zimory Distributed Cloud – Whitepaper. Available online at: http://www.zimory.de/index.php?eID=tx_nawsecuredl&u=0&file=fileadmin/user_upload/pdf/Distributed _Clouds_Whitepaper.pdf&t=1359027268&hash=93c5f42f8c91817a746f7b8cff55fbdc68ae7379. (Accessed on: January 23, 2011).